\documentclass[twocolumn,showpacs,pre,floatfix,superscriptaddress]{revtex4}
\usepackage{color} 
\usepackage{tabularx} 
\usepackage{epsfig}
\usepackage{amsmath} 
\usepackage{amssymb} 
\usepackage{bm}
\usepackage{graphicx}
\usepackage{wasysym}
\usepackage{epsfig}

\begin{document}

\title{Boolean decision problems with competing interactions on
scale-free networks:\\ Equilibrium and nonequilibrium behavior in
an external bias}

\author{Zheng Zhu}
\affiliation {Department of Physics and Astronomy, Texas A\&M University,
College Station, Texas 77843-4242, USA}

\author{Juan Carlos Andresen}
\affiliation{Theoretische Physik, ETH Zurich, CH-8093 Zurich, Switzerland}

\author{M.~A.~Moore}
\affiliation{School of Physics and Astronomy, University of Manchester,
Manchester M13 9PL, UK}

\author{Helmut G. Katzgraber}
\affiliation {Department of Physics and Astronomy, Texas A\&M University,
College Station, Texas 77843-4242, USA}
\affiliation{Materials Science and Engineering Program, Texas A\&M
University, College Station, Texas 77843-3003, USA}

\date{\today}

\begin{abstract}

We study the equilibrium and nonequilibrium properties of Boolean
decision problems with competing interactions on scale-free networks in
an external bias (magnetic field). Previous studies at zero field have
shown a remarkable equilibrium stability of Boolean variables (Ising
spins) with competing interactions (spin glasses) on scale-free
networks. When the exponent that describes the power-law decay of the
connectivity of the network is strictly larger than 3, the system
undergoes a spin-glass transition.  However, when the exponent is equal
to or less than 3, the glass phase is stable for all temperatures.
First, we perform finite-temperature Monte Carlo simulations in a field
to test the robustness of the spin-glass phase and show that the system
has a spin-glass phase in a field, i.e., exhibits a de Almeida--Thouless
line.  Furthermore, we study avalanche distributions when the system is
driven by a field at zero temperature to test if the system displays
self-organized criticality. Numerical results suggest that avalanches
(damage) can spread across the whole system with nonzero probability
when the decay exponent of the interaction degree is less than or equal
to 2, i.e., that Boolean decision problems on scale-free networks with
competing interactions can be fragile when not in thermal equilibrium.

\end{abstract}


\maketitle

\section{Introduction}

Scale-free networks play an integral role in nature, as well as
industrial, technological and sociological applications
\cite{albert:99}.  In these networks, the edge degrees $\{k_i\}$ (the
number of neighbors each node has) are distributed according to a power
law $\lambda$, with the probability $\wp_k$ for a node to have $k$
neighbors given by
\begin{eqnarray}
\label{eq:p_k}
\wp_k &\propto& k^{-\lambda}.
\end{eqnarray}
In the meantime, there have been many studies of Boolean variables on
scale-free networks \cite{bartolozzi:06,herrero:09,lee:06a,weigel:07}
and, more recently, even with competing interactions
\cite{mooij:04,kim:05,ferreira:10,ostilli:11,katzgraber:12}.  There is
general consensus that stable ferromagnetic and spin-glass phases emerge
in these complex systems \cite{katzgraber:12} and that for particular
choices of the decay exponent $\lambda$ the critical temperature
diverges, i.e., Boolean variables with competing interactions are
extremely robust to local perturbations.

However, the behavior of these intriguing systems in an external
magnetic field---which can be interpreted as a global bias---remains to
be fully understood.  Although a replica ansatz works well when
determining the critical temperature of the system
\cite{kim:05,katzgraber:12} in zero field, it is unclear if a stable
spin-glass state persists in a field.  In addition, when studying the
system without local perturbations (i.e., at zero temperature), it is
unclear if ``damage'' in the form of avalanches of Boolean variable
flips triggered by a field can spread easily across the system.

In this work we tackle the two aforementioned problems numerically and
show that at finite temperature Boolean variables with competing
interactions are remarkably robust to global external biases.  In
particular, we show that a de Almeida--Thouless line \cite{almeida:78}
persists to a regime of $\lambda$ where the system is not in the
mean-field Sherrington-Kirkpatrick \cite{sherrington:75} universality
class, i.e., when $\lambda < 4$ \cite{kim:05,katzgraber:12}.

Furthermore, we probe for the existence of self-organized criticality
(SOC) when driving the system at zero temperature with an external
magnetic field across a hysteresis loop. SOC is a property of large
dissipative systems to drive themselves into a scale-invariant critical
state without any special parameter tuning
\cite{newman:94,cieplak:94,schenk:02,pazmandi:99,goncalves:08}. It is a
phenomenon found in many problems ranging from earthquake statistics to
the structure of galaxy clusters. As such, studying SOC on scale-free
networks might help us gain a deeper understanding on how avalanches,
i.e., large-scale perturbations, might spread across scale-free networks
that are so omnipresent in nature.  Recent simulations
\cite{andresen:13} have shown that a diverging number of neighbors is
the key ingredient to obtain SOC in glassy spin systems. In scale-free
graphs the average edge degree diverges if $\lambda \le 2$.  As such, it
might be conceivable that in this regime spin glasses on scale-free
graphs exhibit SOC.  However, it is unclear what happens for $\lambda >
2$ where the number of neighbors each spin has is finite in the
thermodynamic limit, or how the fraction of ferromagnetic versus
antiferromagnetic bonds influences the scaling of the avalanche
distributions.  Within the spin-glass phase, for Gaussian disorder and
bimodal disorder with the same fraction $p$ of ferromagnetic and
antiferromagnetic bonds, we find that when $\lambda \le 2$ Boolean
variables with competing interactions always display SOC like the
mean-field Sherrington-Kirkpatrick model \cite{pazmandi:99}.  For
$\lambda > 2$ and with bimodal disorder, a critical line in the
$p$--$\lambda$ plane emerges along which perturbations to the system are
scale free, but not self-organized critical because the fraction of
ferromagnetic bonds has to be carefully tuned. The latter is reminiscent
of the behavior found in the random-field Ising model
\cite{sethna:93,perkovic:95,perkovic:99,kuntz:98,sethna:04}, as well as
random-bond \cite{vives:94} and random-anisotropy Ising models
\cite{vives:01}.

The paper is structured as follows. Section \ref{sec:model} introduces
the Hamiltonian studied, followed by numerical details, observables, and
results from equilibrium Monte Carlo simulations in
Sec.~\ref{sec:equilibrium}. Section \ref{sec:Non-equilibrium} presents
our results on nonequilibrium avalanches on scale-free graphs, followed
by concluding remarks. In the appendix we outline our analytical
calculations to determine the de Almeida--Thouless for spin glasses on
scale-free graphs.

\section{Model}
\label{sec:model}

The Hamiltonian of the Edwards-Anderson Ising spin glass on a scale-free
graph in an external magnetic field is given by
\begin{equation}
{\mathcal H}(\{s_i\})
    =-\sum_{i<j}^N J_{ij} \varepsilon_{ij} s_i \,s_j-\sum_i H_is_i , 
\label{eq:ham}
\end{equation}
where the Ising spins $s_i \in\{\pm 1\}$ lie on the vertices of a
scale-free graph with $N$ sites and the interactions are given by
\begin{eqnarray}
\label{eq:bond_disorder}
{\mathcal P}(J_{ij}, \varepsilon_{ij})
=\wp_J(J_{ij}) 
\left[  \Big(1-\frac{K}{N} \Big) \delta( \varepsilon_{ij})
+ 
\frac{K}{N} \delta( \varepsilon_{ij}-1) \right]. 
\end{eqnarray}
If a bond is present, we set $\varepsilon_{ij}=1$, otherwise
$\varepsilon_{ij}=0$.  $K$ represents the mean connectivity of the
scale-free graph.  The connectivity of site $i$, $k_i:=\sum_j
\varepsilon_{ij}$, is sampled from a scale-free distribution as done in
Ref.~\cite{katzgraber:12}.  The interactions between the spins $J_{ij}$
are independent random variables drawn from a Gaussian distribution with
zero mean and standard deviation unity, i.e.,
\begin{equation}
\wp_J(J_{ij}) \sim \exp{(-J_{ij}^2/2)}  \, .
\label{eq:gauss}
\end{equation}
In the nonequilibrium studies we also study bimodal-distributed disorder
where we can change the fraction of ferromagnetic bonds $p$, i.e.,
\begin{equation}
\wp_J(J_{ij}) =  p\delta(J_{ij} + 1) + (1-p)\delta(J_{ij} - 1) \, .
\label{eq:bim}
\end{equation}
Finally, for the finite-temperature studies we use random fields drawn
from a Gaussian distribution with zero mean and standard deviation $H_r$
in Eq.~\eqref{eq:ham}, instead of a uniform field. This allows us to
perform a detailed equilibration test of the Monte Carlo method
\cite{katzgraber:01,katzgraber:09b}.

The scale-free graphs are generated using preferential attachment with
slight modifications \cite{barabasi:99}. Details of the method are described
in Ref.~\cite{katzgraber:12}. We impose an upper bound on the
allowed edge degrees, $k_\mathrm{max} = \sqrt{N}$.  Although we can, in
principle, generate graphs with $k$ exceeding $\sqrt{N}$, the ensemble
is poorly defined in this case: Even randomly chosen graphs cannot be
uncorrelated \cite{burda:03,boguna:04,catanzaro:05}.  Furthermore, to
prevent dangling ends that do not contribute to frustrated loops in the
system, we set a lower bound to the edge degree, namely
$k_\mathrm{min}=3$.

\section{Equilibrium properties in a field}
\label{sec:equilibrium}

In equilibrium, the behavior of spin glasses in a magnetic field is
controversial
\cite{young:04,katzgraber:05c,joerg:08a,katzgraber:09b,banos:12,baity:13}.
While the infinite-range (mean-field) Sherrington-Kirkpatrick (SK) model
\cite{sherrington:75} has a line of transitions at finite field known as
the de Almeida--Thouless (AT) line \cite{almeida:78} that separates the
spin-glass phase from the paramagnetic phase at finite
fields or temperatures, it has not been definitely established whether an
AT line occurs in systems with short-range interactions.  Spin glasses
on scale-free networks are somewhat ``in between'' the infinite-range
and short-range limits depending on the exponent $\lambda$.  As such, it
is unclear if a spin-glass state will persist when an external field $H$
is applied, especially when the spin-glass transition at zero field
occurs at finite temperatures, i.e., for $\lambda > 3$.

Note that spin glasses on scale-free graphs share the same universality
class as the SK model if $\lambda > 4$ \cite{katzgraber:12}. As such, in
this regime, one can expect an AT line. However, for $3 < \lambda < 4$,
where $T_c < \infty$, the critical exponents depend on the exponent
$\lambda$ \cite{kim:05,katzgraber:12}. Therefore, it is unclear if a
spin-glass state in a field will persist. For $\lambda \le 3$ the
critical temperature diverges with the system size, i.e., we also expect
the system to have a spin-glass state for finite fields. We therefore
focus on two values of $\lambda$, namely $\lambda = 4.50$ (deep within
the SK-like regime because $\lambda = 4$ has logarithmic corrections)
\cite{katzgraber:12} and $\lambda = 3.75$ (where the existence of an AT
line remains to be determined).

\subsection{Observables}

In simulations, it is most desirable to perform a finite-size scaling
(FSS) of dimensionless quantities. One such quantity, the Binder
ratio \cite{binder:81}, turns out to be poorly behaved in an external
field in short-range systems \cite{ciria:93b}. Therefore, to determine
the location of a spin-glass phase transition we measure the connected
spin-glass susceptibility given by
\begin{equation}
\chi = \frac{1}{N}
	\displaystyle\sum_{i,j}[\left(\langle{s_is_j}\rangle_T -
	\langle{s_i}\rangle_T\langle{s_j}\rangle_T\right)^2]_{\rm av},
\label{:chidef}
\end{equation}
where $\langle\cdots\rangle_T$ denotes a thermal average and
$[\cdots]_{\rm av}$ an average over both the bond disorder and different
network instances. $N$ is the number of spins.  To avoid bias, each
thermal average is obtained from separate copies (replicas) of the
spins. This means that we simulate four independent replicas at each
temperature.

For any spin glass outside the mean-field regime, the scaling behavior
of the susceptibility is given by \cite{katzgraber:12}
\begin{equation}
\chi = N^{2 - \eta} \widetilde{C}
\left(N^{1/\nu}[\beta - \beta_c]\right) \, ,
\label{eq:scalchi}
\end{equation}
where $\nu$ and $\eta$ are the correlation length and susceptibility
exponents, respectively, and $\beta_c=1/T_c$ is the inverse temperature
for a given field strength $H_r$.  

For $\lambda < 4$ (see the appendix for details) we expect the critical
exponent $\gamma = 1$. This is only possible if $2 - \eta = 1/\nu$ in
Eq.~\eqref{eq:scalchi}.  Using the standard scaling relation $\alpha + 2
\beta + \gamma = 2$, the hyperscaling relation $d \nu= 2 - \alpha$
(which we assume will hold when $\lambda<4$), and allowing for the
nonstandard meaning of $\nu$ in this paper (it is equal to $d \nu$ in
standard notation where $d$ is here the dimensionality of the system),
it follows for $\lambda <4$, where $\beta=1/(\lambda-3)$ (see the
appendix and Ref.~\cite{kim:05}) that
\begin{equation}
\nu = \frac{\lambda - 1}{\lambda - 3} 
\;\;\;\;{\rm and} \;\;\;\;
\eta = 2 - \frac{1}{\nu}.
\label{eq:exp}
\end{equation}
For the case of $\lambda = 3.75$ this means that $\nu = 11/3$ and
therefore $\eta = 2 - 1/\nu = 19/11$. As such, curves of $\chi/N^{3/11}$
should have the same scaling behavior as the Binder ratio.

For $\lambda > 4$, the finite-size scaling form presented in
Eq.~\eqref{eq:scalchi} is replaced by \cite{katzgraber:09b,larson:13}
\begin{equation}
\chi = N^{1/3} \widetilde{C}
\left(N^{1/3}[\beta - \beta_c]\right) \, .
\label{eq:scaldefm}
\end{equation}
In this case the scaling is simpler because the exponents are fixed and
independent of $\lambda$, i.e., $1/\nu = 2 - \eta = 1/3$. Here, curves
of $\chi/N^{1/3}$ should have the same scaling behavior as the Binder
ratio.  Performing a finite-size scaling of the data therefore allows
one to detect the transition to high precision.

Finally, note that the aforementioned study is, strictly speaking, only
valid at zero field.  Although $\gamma = 1$ across the AT line, there is
no explicit calculation of the critical exponent $\beta$ in a field.
While our data suggest that the values of the zero-field exponents might
be the same as those for finite external fields,  the accuracy of our
results for the exponents in a field is limited by large finite-size
corrections.

\subsection{Equilibration scheme and simulation parameters}

The simulations are done using the parallel tempering Monte Carlo
method \cite{geyer:91,hukushima:96}. The spins couple to site-dependent
random fields $H_i$ chosen from a Gaussian distribution with zero mean
$[H_i]_{\rm av}=0$ and standard deviation $[H_i^2]_{\rm av}^{1/2}=H_r$.
Simulations are performed at zero field as well as at $H_r = 0.1$,
$0.2$, $0.3$, and $0.4$. Using Gaussian disorder, we can use a
strong equilibration test to ensure that the data are in thermal
equilibrium \cite{katzgraber:01,katzgraber:09b,katzgraber:12}.
Here, the internal energy per spin
\begin{equation}
U = (1/N) [\langle {\mathcal H}\rangle_T]_{\rm av} \, ,
\end{equation}
with ${\mathcal H}$ defined in Eq.~\eqref{eq:ham}, has to equate an
expression derived from both the link overlap $q_4$ given by
\begin{equation}
q_4 =  \frac{1}{N_b}  \sum_{i,j} \varepsilon_{ij}
        s_i^\alpha s_j^\alpha s_i^\beta s_j^\beta \, ,
\end{equation}
and the spin overlap
\begin{equation}
q =  \frac{1}{N_b}  \sum_{i} s_i^\alpha s_i^\beta \, .
\end{equation}
Here $\alpha$ and $\beta$ represent two copies of the system with the
same disorder and $N_b$ represents the number of neighbors each spin has
for a given sample (graph instance). Note that because in
Eq.~\eqref{:chidef} we already simulate four replicas, we actually
perform an average over all four-replica permutations.

\begin{figure}
\includegraphics[width=\columnwidth]{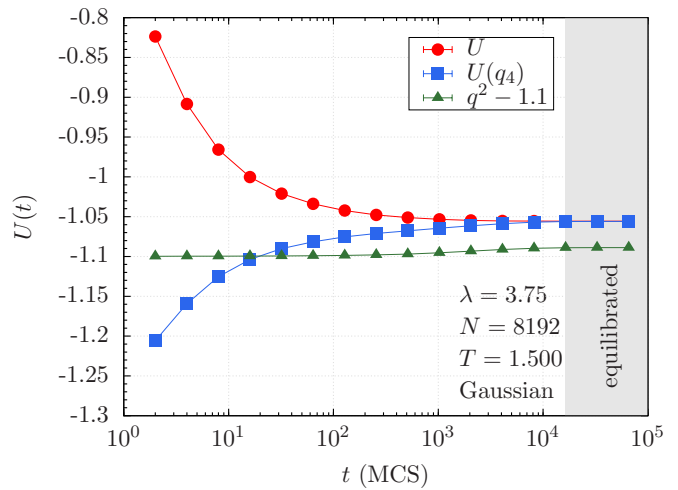}
\caption{(Color online)
Equilibration test for $N = 8192$ spins at $T = 1.500$ (lowest
temperature simulated) and $\lambda = 3.75$.  Once the data for the
energy $U$ and the energy computed from $q_4$ [$U(q_4)$] agree, the
system is in thermal equilibrium (shaded area).  At this point data for
$q^2$ are also independent of Monte Carlo time.  Note that the data for
$q^2$ are shifted by a constant factor of $1.1$ for better comparison.
Error bars are smaller than the symbols.
}
\label{fig:equil}
\end{figure}

The system is in thermal equilibrium if 
\begin{equation}
U=U(q_4)= - \frac{1}{T}\,
	\left[\left\langle 
		\frac{N_b}{N}\, (1 - q_4) + H_r^2\,(1-q)
	\right\rangle\right]_{\rm av} \, .
\label{:eqa}
\end{equation}
Sample data are shown in Fig.~\ref{fig:equil}.  The energy $U$ computed
directly is compared to the energy computed from the link overlap
$U(q_4)$. The data for both quantities approach a limiting value from
opposite directions. Once $U = U(q_4)$, the data for $q^2$ (shifted for
better viewing in Fig.~\ref{fig:equil}) are also in thermal equilibrium.
The simulation parameters are shown in Table \ref{tab:atlineparams}.

\begin{table}
\caption{
Parameters of the simulation: For each exponent $\lambda$ and system
size $N$, we compute $N_{\rm sa}$ disorder or network instances. $N_{\rm
sw} = 2^b$ is the total number of Monte Carlo sweeps for each of the
$4N_T$ replicas for a single instance, $T_{\rm min}$ [$T_{\rm max}$] is
the lowest [highest] temperature simulated, and $N_T$ is the number of
temperatures used in the parallel tempering method for each system size
$N$.
\label{tab:atlineparams}
}
\begin{tabular*}{\columnwidth}{@{\extracolsep{\fill}} l r r r r r r r  }
\hline
\hline
$\lambda$ &$H_r$ & $N$ & $N_{\rm sa}$ & $b$  & $T_{\rm min}$ &$T_{\rm max}$ & $N_T$ \\
\hline
$3.75$ & $0.0$ & $2048$  & $9600$  & $16$ & $1.5000$  & $3.0000$  & $30$   \\
$3.75$ & $0.0$ & $3072$  & $9600$  & $16$ & $1.5000$  & $3.0000$  & $30$   \\
$3.75$ & $0.0$ & $4096$  & $9600$  & $16$ & $1.5000$  & $3.0000$  & $30$   \\
$3.75$ & $0.0$ & $6144$  & $9600$  & $16$ & $1.5000$  & $3.0000$  & $30$   \\
$3.75$ & $0.0$ & $8192$  & $9600$  & $16$ & $1.5000$  & $3.0000$  & $30$   \\
\hline
$3.75$ & $0.1$ & $512$   & $9600$  & $17$ & $0.9000$  & $3.0000$  & $50$   \\
$3.75$ & $0.1$ & $768$   & $9600$  & $17$ & $0.9000$  & $3.0000$  & $50$   \\
$3.75$ & $0.1$ & $1024$  & $9600$  & $17$ & $0.9000$  & $3.0000$  & $50$   \\
$3.75$ & $0.1$ & $1536$  & $9600$  & $18$ & $0.9000$  & $3.0000$  & $50$   \\
$3.75$ & $0.1$ & $2048$  & $2400$  & $18$ & $0.9000$  & $3.0000$  & $50$   \\
\hline
$3.75$ & $0.2$ & $768$   & $9600$  & $17$ & $0.9000$  & $3.0000$  & $50$   \\
$3.75$ & $0.2$ & $1024$  & $9600$  & $17$ & $0.9000$  & $3.0000$  & $50$   \\
$3.75$ & $0.2$ & $1536$  & $9600$  & $18$ & $0.9000$  & $3.0000$  & $50$   \\
$3.75$ & $0.2$ & $2048$  & $2400$  & $18$ & $0.9000$  & $3.0000$  & $50$   \\
$3.75$ & $0.2$ & $4096$  & $2400$  & $19$ & $0.9000$  & $3.0000$  & $50$   \\
\hline
$3.75$ & $0.3$ & $256$   & $9600$  & $17$ & $0.9000$  & $3.0000$  & $50$   \\
$3.75$ & $0.3$ & $512$   & $9600$  & $18$ & $0.9000$  & $3.0000$  & $50$   \\
$3.75$ & $0.3$ & $1024$  & $9600$  & $18$ & $0.9000$  & $3.0000$  & $50$   \\
$3.75$ & $0.3$ & $2048$  & $2400$  & $18$ & $0.9000$  & $3.0000$  & $50$   \\
\hline
$3.75$ & $0.4$ & $256$   & $9600$  & $18$ & $0.9000$  & $3.0000$  & $50$   \\
$3.75$ & $0.4$ & $512$   & $9600$  & $18$ & $0.9000$  & $3.0000$  & $50$   \\
$3.75$ & $0.4$ & $1024$  & $9600$  & $18$ & $0.9000$  & $3.0000$  & $50$   \\
$3.75$ & $0.4$ & $2048$  & $2400$  & $18$ & $0.9000$  & $3.0000$  & $50$   \\
\hline
$4.50$ & $0.0$ & $1024$   & $9600$  & $16$ & $1.0000$  & $3.0000$  & $30$   \\
$4.50$ & $0.0$ & $2048$   & $9600$  & $16$ & $1.0000$  & $3.0000$  & $30$   \\
$4.50$ & $0.0$ & $4096$   & $9600$  & $16$ & $1.0000$  & $3.0000$  & $30$   \\
$4.50$ & $0.0$ & $8192$   & $9600$  & $16$ & $1.0000$  & $3.0000$  & $30$   \\
\hline
$4.50$ & $0.1$ & $512$    & $9600$  & $17$ & $0.9000$  & $3.0000$  & $50$   \\
$4.50$ & $0.1$ & $1024$   & $9600$  & $17$ & $0.9000$  & $3.0000$  & $50$   \\
$4.50$ & $0.1$ & $2048$   & $9600$  & $18$ & $0.9000$  & $3.0000$  & $50$   \\
$4.50$ & $0.1$ & $4096$   & $2400$  & $18$ & $0.9000$  & $3.0000$  & $50$   \\
\hline
$4.50$ & $0.2$ & $256$    & $9600$  & $18$ & $0.6000$  & $3.0000$  & $50$   \\
$4.50$ & $0.2$ & $512$    & $9600$  & $18$ & $0.6000$  & $3.0000$  & $50$   \\
$4.50$ & $0.2$ & $1024$   & $9600$  & $18$ & $0.6000$  & $3.0000$  & $50$   \\
$4.50$ & $0.2$ & $2048$   & $2400$  & $19$ & $0.6000$  & $3.0000$  & $50$   \\
\hline
$4.50$ & $0.3$ & $64$     & $9600$  & $18$ & $0.3000$  & $3.0000$  & $50$   \\
$4.50$ & $0.3$ & $128$    & $9600$  & $19$ & $0.3000$  & $3.0000$  & $50$   \\
$4.50$ & $0.3$ & $256$    & $9600$  & $20$ & $0.3000$  & $3.0000$  & $50$   \\
$4.50$ & $0.3$ & $512$    & $9600$  & $22$ & $0.3000$  & $3.0000$  & $50$   \\
\hline
$4.50$ & $0.4$ & $90$     & $9600$  & $19$ & $0.3000$  & $3.0000$  & $50$   \\
$4.50$ & $0.4$ & $128$    & $9600$  & $19$ & $0.3000$  & $3.0000$  & $50$   \\
$4.50$ & $0.4$ & $180$    & $9600$  & $19$ & $0.3000$  & $3.0000$  & $50$   \\
$4.50$ & $0.4$ & $256$    & $9600$  & $20$ & $0.3000$  & $3.0000$  & $50$   \\
\hline
\hline
\end{tabular*}
\end{table}

\subsection{Numerical results for $\lambda = 4.50$}

\begin{figure}
\includegraphics[width=\columnwidth]{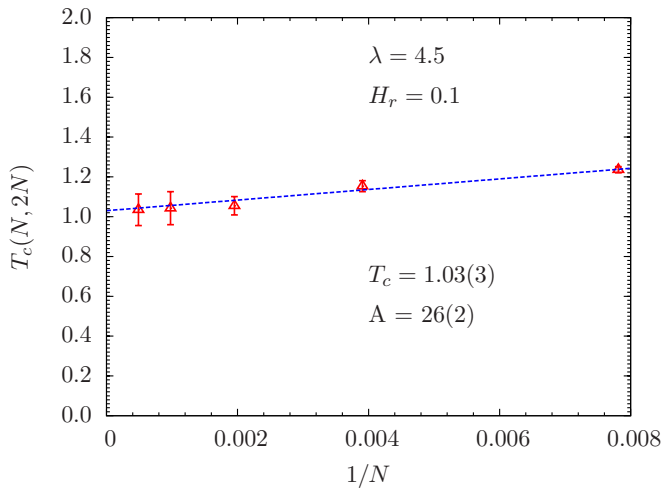}
\caption{(Color online)
Extrapolation to the thermodynamic limit for the critical temperature
$T_c$ for $\lambda = 4.50$ and $H_r = 0.1$. We determine the crossing
points of critical temperatures of the susceptibility expression for
pairs of system sizes $N$ and $2N$. Using Eq.~\eqref{eq:fit} with
$\omega = 1$ we extrapolate the data to the thermodynamic limit.  This
allows us to take into account corrections to scaling in an unbiased
way.
}
\label{fig:4.5_0.1-ex}
\end{figure}

Corrections to scaling are large for this model despite the large
system sizes and number of samples studied.  As previously stated, we
expect that for $\lambda = 4.50$ a spin-glass state is stable towards an
external field because for $\lambda > 4$ the system shares the same
universality class as the SK model. To determine the AT line, we plot
$\chi/N^{1/3}$ versus the inverse temperature $\beta = 1/T$. Because
$\chi/N^{1/3}$ is a dimensionless function [see
Eq.~\eqref{eq:scaldefm}], data for different system sizes should cross
at the putative field-dependent transition temperature.  To cope with
corrections to scaling and obtain a precise estimate of the critical
temperature, we study the crossing temperatures $T_c(N,2N)$ for pairs of
system sizes $N$ and $2N$ assuming
\begin{equation}
T_c(N,2N) = T_c + A/N^{\omega} \, ,
\label{eq:fit}
\end{equation}
where $A$ is a fitting parameter and empirically $\omega = 1$.  An
example extrapolation is shown in Fig.~\ref{fig:4.5_0.1-ex} for $\lambda
= 4.50$ and $H_r = 0.1$. A linear fit is very stable and the
extrapolation to the thermodynamic limit clear. Statistical error bars
are determined via a bootstrap analysis \cite{katzgraber:06} using the
following procedure: For each system size $N$ and $N_{\rm sa}$ disorder
realizations, a randomly selected bootstrap sample of $N_{\rm sa}$
disorder realizations is generated. With this random sample, an estimate
of $\chi/{N^{1/3}}$ is computed for each temperature. The crossing
temperature for pairs of $N$ and $2N$ is obtained by fitting the data to
a third-order polynomial and a subsequent root determination.  We repeat
this procedure $N_{\rm boot} = 500$ times for each lattice size and then
assemble $N_{\rm boot}$ complete data sets (each having results for
every system size $N$) by combining the $i$th bootstrap sample for each
size for $i = 1$, $\ldots$, $N_{\rm boot}$.  The nonlinear fit to
Eq.~\eqref{eq:fit} is then carried out on each of these $N_{\rm boot}$
sets, thus obtaining $N_{\rm boot}$ estimates of the fit parameters
$T_c$ and $A$. Because the bootstrap sampling is done with respect to
the disorder realizations which are statistically independent, we can
use a conventional bootstrap analysis to estimate statistical error bars
on the fit parameters. These are comparable to the standard deviation
among the $N_{\rm boot}$ bootstrap estimates.

\begin{figure}
\includegraphics[width=\columnwidth]{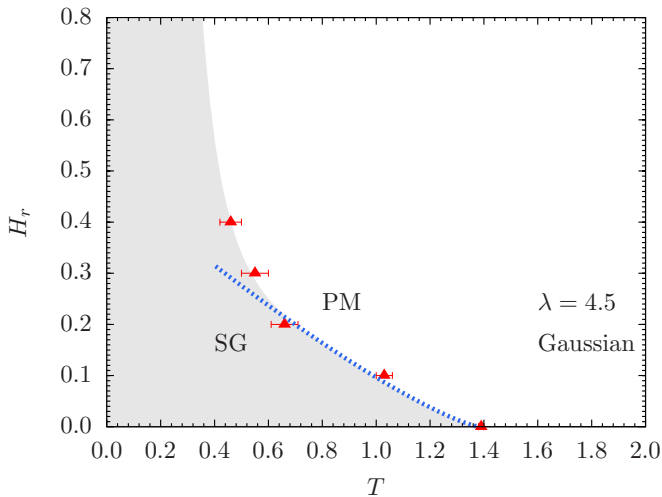}
\caption{(Color online)
Field $H_r$ versus temperature $T$ phase diagram for an Ising spin glass on
a scale-free graph with $\lambda = 4.50$. The data points separate a
paramagnetic (PM) from a spin-glass (SG) state. The shaded area
is intended as a guide to the eye. The dotted (blue) line is a calculation
of the AT line in the $H_r \to 0$ limit.
}
\label{fig:tc_4.5}
\end{figure}

The obtained estimates of $T_c$ are listed in Table
\ref{tab:critparams}. Figure \ref{fig:tc_4.5} shows the
field--temperature phase diagram for $\lambda = 4.50$. The shaded area
is intended as a guide to the eye. The critical line separates a
paramagnetic (PM) from a spin-glass (SG) phase.  The dotted (blue) line
represents the AT line computed analytically (appendix) in the limit of
$H_r \to 0$. For $4 < \lambda < 5$ the shape of the AT line is given by
Eq.~\eqref{eqn45}. The analytical approximation fits the data for
$\lambda = 4.5$ very well with $H_r(T) \sim C_{4.5} (1 - T/T_c)^{5/4}$
and $C_{4.5} = 0.48(3)$.

\subsection{Numerical results for $\lambda = 3.75$}

Because for $\lambda < 4$ we are no longer in the SK universality class,
it is {\em a priori} unclear if a spin-glass state in a field will exist.
Furthermore, when $\lambda = 3.75$, a finite-size scaling according to
Eq.~\eqref{eq:scalchi} has to be performed. Because it is not possible
to define a distance metric on a scale-free network, there is no notion
of a correlation length or spin-spin correlation function. As such, the
critical exponents $\nu$ (that describes the divergence of the
correlation length) and $\eta$ (also known as the anomalous dimension)
have to be treated carefully. However, we will assume that
Eq.~\eqref{eq:scalchi} is valid in this regime on generic finite-size
scaling grounds and treat $\nu$ and $\eta$ as parameters when $H_r > 0$
with no special meaning attached to them. In addition, we fix $\nu =
11/3$ and $\eta = 2 - 1/\nu$ --- the zero-field values of the critical
exponents --- and scale the data at finite fields assuming these
exponents are valid also when $H_r > 0$.

To determine $T_c(H_r)$, we perform a finite-size scaling analysis of
the susceptibility data according to Eq.~\eqref{eq:scalchi}. To
determine the optimal value of $T_c = 1/\beta_c$ that scales the data
best we use the approach developed in Ref.~\cite{katzgraber:06}.  We
assume that the scaling function in Eq.~\eqref{eq:scalchi} can be
represented by a third-order polynomial $y(x) = c_0 + c_1x + c_2x^2 +
c_3x^3$ for $|x| \lesssim 1$ and do a global fit to the seven parameters
$c_i$ with $i \in \{0, \ldots, 3\}$, $\beta_c$, $\eta$, and $\nu$.  Here
$y = \chi/{N^{2-\eta}}$ and $x=N^{1/\nu}[\beta - \beta_c]$.  After
performing a Levenberg-Marquardt minimization combined with a bootstrap
analysis we determine the optimal critical parameters with an unbiased
statistical error bar.

\begin{figure}

\includegraphics[width=\columnwidth]{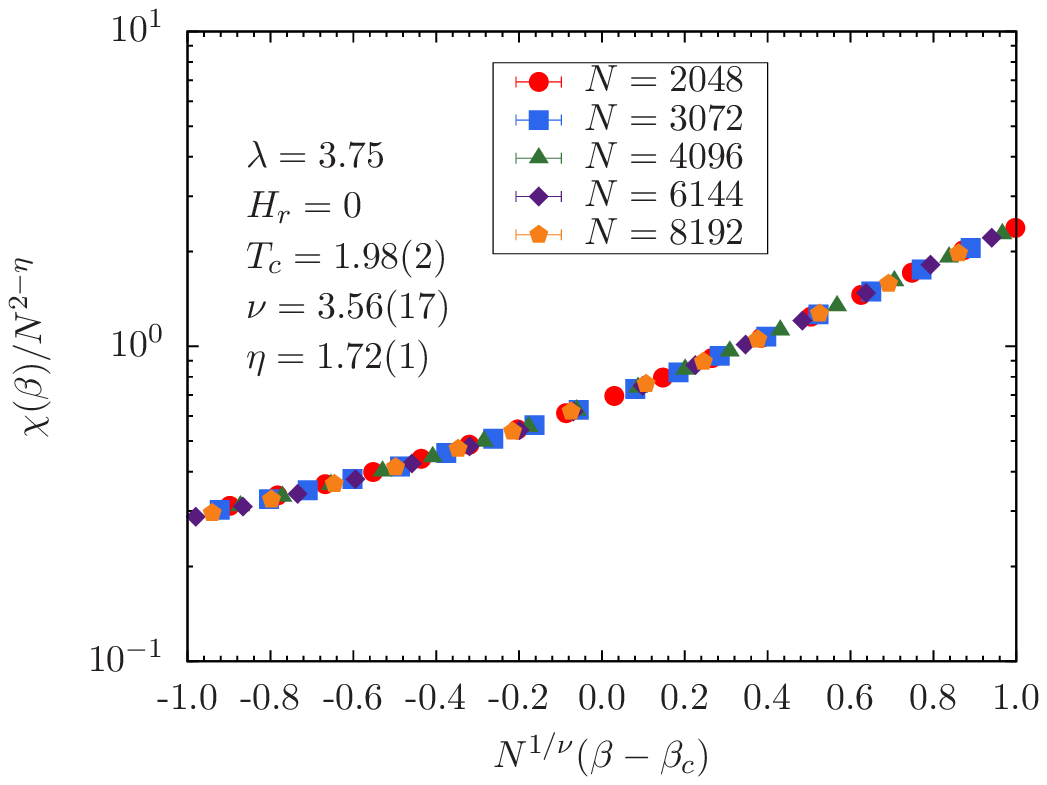}

\vspace*{1.0em}

\includegraphics[width=\columnwidth]{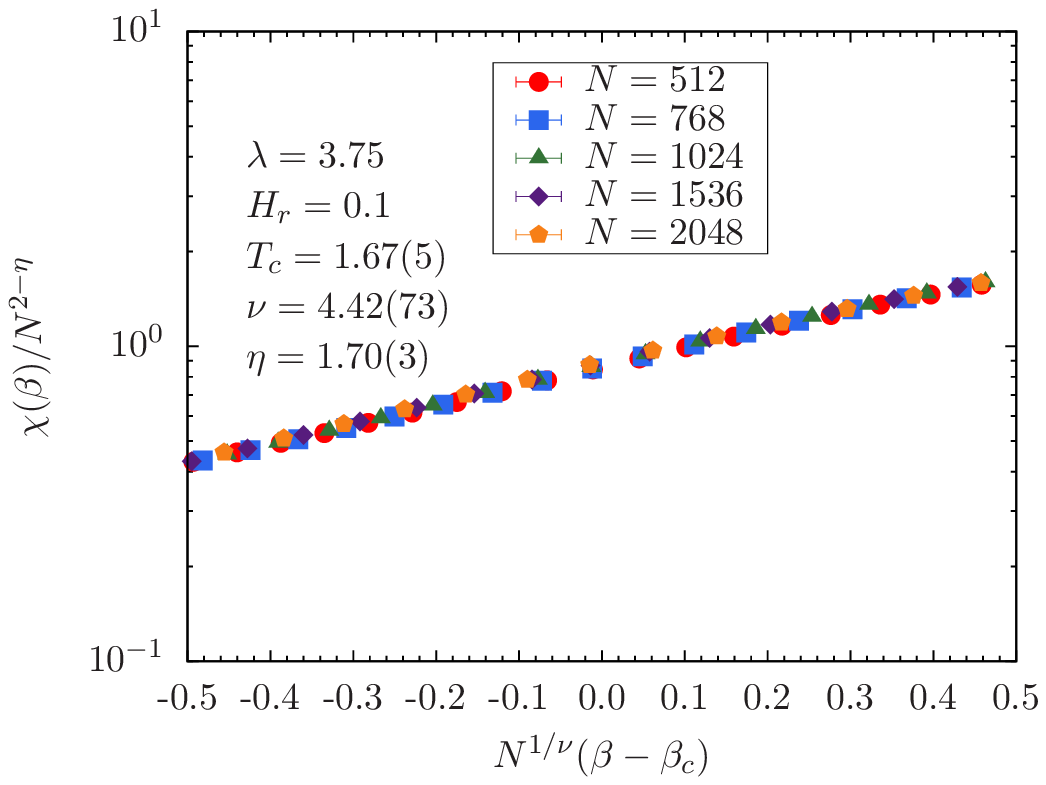}

\caption{(Color online)
Finite-size scaling analysis of $\chi/{N^{2-\eta}}$ as a function of
$N^{1/\nu}(\beta - \beta_c)$ for an Ising spin glass on a scale-free
network with Gaussian disorder and $\lambda = 3.75$. The data at zero
field (top panel) scale very well. The bottom panel shows representative
data for $H_r = 0.1$ scaled according to Eq.~\eqref{eq:scalchi}.
Error bars are smaller than the symbols.
}
\label{fig:ss_collapse_3.75}
\end{figure}

\begin{figure}
\includegraphics[width=\columnwidth]{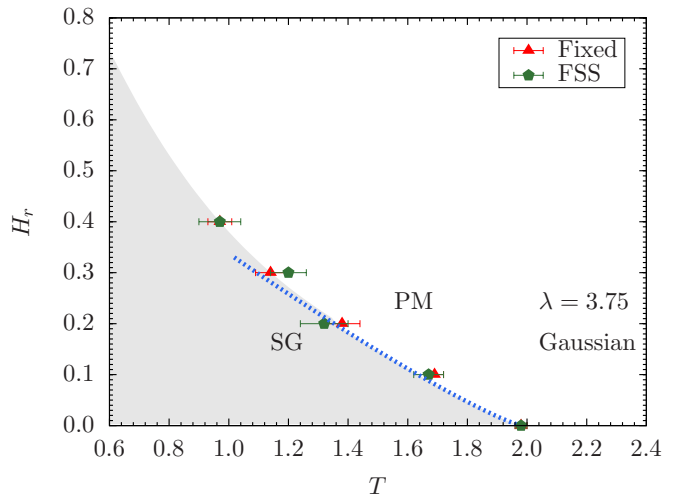}
\caption{(Color online)
Field $H_r$ versus temperature $T$ phase diagram for an Ising spin glass on
a scale-free graph with $\lambda = 3.75$. The data points separate a
paramagnetic (PM) from a spin-glass (SG) state. The shaded area is
intended as a guide to the eye.  The dotted (blue) line is a calculation
of the AT line in the $H_r \to 0$ limit.  Note that estimates for the
critical temperature $T_c$ from a finite-size scaling analysis (FSS)
according to Eq.~\eqref{eq:scalchi} with $T_c$, $\eta$, and $\nu$ as
free parameters agree within error bars with estimates at finite fields
where $\eta = 19/11$ and $\nu = 11/3$ are used as fixed parameters
(labeled with ``Fixed'' in the plot).
}
\label{fig:tc_3.75}
\end{figure}

Figure \ref{fig:ss_collapse_3.75} shows two representative scaling
collapses at zero and nonzero field values. The data scale well and
allow one to determine the critical temperature with good precision
despite the difficulties that scaling the spin-glass susceptibility
poses \cite{katzgraber:06}. Note that for zero field we obtain $\eta =
1.72(1)$ and $\nu = 3.56(17)$, which agree very well with the analytical
expressions $\eta = 19/11 = 1.72\ldots$ and $\nu = 11/3 = 3.66\ldots$.
However, for finite fields deviations are visible.  A summary of the
relevant fitting parameters is listed in Table \ref{tab:critparams}.
Note that the value of $\eta$ for different fields agrees within error
bars. However, fluctuations are larger for $\nu$. One can expect that
the universality class of the system does not change along the AT line
\cite{binder:86}. Therefore, and because it is hard to simulate large
systems for large fields, we also determine $T_c$ by fixing $\eta =
19/11$ and $\nu = 11/3$.  As listed in Table \ref{tab:critparams}, both
estimates agree within error bars. This is also visible in
Fig.~\ref{fig:tc_3.75} which shows the AT line for $\lambda = 3.75$.
Overall, the analysis using the zero-field estimates for $\eta$ and
$\nu$ gives more accurate results.  The dotted (blue) line in
Fig.~\ref{fig:tc_3.75} is our analytical estimate of the AT line
computed in the $H_r \to 0$ limit (appendix). The estimate fits the data
well with $H_r(T) \sim C_{3.75} (1 - T/T_c)^{7/6}$ and $C_{3.75} =
0.76(5)$.

\begin{table}
\caption{
Critical parameters $T_c$, $\nu$, and $\eta$ for a spin glass with
Gaussian random bonds defined on a scale-free graph. The data for
$\lambda = 4.50$ have been determined using the mean-field finite-size
scaling expression in Eq.~\eqref{eq:scaldefm}. In this case one can, in
principle, define $\eta = 5/3$ and $\nu = 3$, although these should be
viewed as parameters placed in Eq.~\eqref{eq:scalchi} to obtain
Eq.~\eqref{eq:scaldefm}. For $\lambda = 3.75$ we determine the critical
parameters using Eq.~\eqref{eq:scalchi}. The starred estimates of $T_c$
for $H_r > 0$ have been determined by using the zero-field estimates of
$\eta = 19/11$ and $\nu = 11/3$ as fixed.  Both $T_c$ and $T_c^\star$
agree within error bars, except statistical fluctuations are smaller for
$T_c^\star$ because there are fewer fitting parameters.
\label{tab:critparams}
}
\begin{tabular*}{\columnwidth}{@{\extracolsep{\fill}} l l l l l r  }
\hline
\hline
$\lambda$ &$H_r$ & $T_c$   & $T_c^\star$ & $\nu$ & $\eta$ \\
\hline
$3.75$ & $0.0$ & $1.98(2)$ & $1.97(1)$ & $3.56(17)$ & $1.72(1)$ \\
$3.75$ & $0.1$ & $1.67(5)$ & $1.68(3)$ & $4.42(73)$ & $1.70(3)$ \\
$3.75$ & $0.2$ & $1.32(8)$ & $1.39(5)$ & $6.53(61)$ & $1.72(2)$ \\
$3.75$ & $0.3$ & $1.20(6)$ & $1.16(4)$ & $3.31(32)$ & $1.74(2)$ \\
$3.75$ & $0.4$ & $0.97(7)$ & $1.00(4)$ & $3.68(46)$ & $1.72(2)$ \\[2mm]
$4.50$ & $0.0$ & $1.39(1)$ &           & $3$        & $5/3$     \\
$4.50$ & $0.1$ & $1.03(3)$ &           & $3$        & $5/3$     \\
$4.50$ & $0.2$ & $0.66(5)$ &           & $3$        & $5/3$     \\
$4.50$ & $0.3$ & $0.55(5)$ &           & $3$        & $5/3$     \\
$4.50$ & $0.4$ & $0.46(4)$ &           & $3$        & $5/3$     \\
\hline
\hline
\end{tabular*}
\end{table}

\section{Nonequilibrium properties in a field}
\label{sec:Non-equilibrium}

It has recently been shown that a key ingredient for the existence of
SOC in glassy spin systems is a diverging number of neighbors
\cite{andresen:13}.  Scale-free networks have a power-law degree
distribution. If the exponent $\lambda \le 2$, then scale-free networks
have an average number of neighbors $K$ that diverges with the system
size. Therefore, it is possible that SOC might be present in this
regime.  To test this prediction, in this section we compute
nonequilibrium avalanche distributions of spin flips driven by an
external field.

\subsection{Numerical details and measured observables}

We study the Hamiltonian in Eq.~\eqref{eq:ham} either with Gaussian
[Eq.~\eqref{eq:gauss}] or bimodal [Eq.~\eqref{eq:bim}] disorder.  The
external magnetic field used to drive the avalanches is uniform rather
than drawn from a Gaussian distribution, i.e., $H_i = H$ in
Eq.~\eqref{eq:ham}.  Spin-flip avalanches are triggered by using
zero-temperature Glauber dynamics
\cite{sethna:93,perkovic:99,katzgraber:02b,andresen:13}. In this
approach one computes the local fields 
\begin{equation}
h_i = \sum_j{J_{ij}}{S_j} -H
\end{equation}
felt by each spin.  A spin is unstable if the stability $h_iS_i < 0$ is
negative.  The initial field $H$ is selected to be larger than the
largest local field, i.e., $H > |h_i|$ $\forall i$. Furthermore, we set
all spins $S_i = +1$.  The spins are then sorted by local fields and the
field $H$ reduced until the stability of the first sorted spin is
negative, therefore making the spin unstable. This (unstable) spin is
flipped, then the local fields of all other spins updated, and the most
unstable spin is flipped again until all spins are stable, i.e., the
avalanche ends.  Simulation parameters are shown in Table
\ref{tab:socparams}.

\begin{table}
\caption{
Simulation parameters in the nonequilibrium study with both Gaussian and
bimodal-distributed random bonds: For each exponent $\lambda$ we study
systems of $N=500 \times 2^m$ spins with $m \in \{1, \ldots, m_{\rm
max}\}$. For Gaussian disorder, when $\lambda<4$, we also simulate
systems with $48\,000$ spins ($m = 6$ corresponds to $32\,000$ spins).
All distributions are computed using $N_{\rm sa}$ disorder realizations.
\label{tab:socparams}
}
\begin{tabular*}{\columnwidth}{@{\extracolsep{\fill}} l l l r }
\hline
\hline
disorder type  & $\lambda$ & $m_{\rm max}$ & $N_{sa}$ \\
\hline
Gaussian & $1.50$ & $6$   & $12\,000$  \\
Gaussian & $2.00$ & $6$   & $12\,000$  \\
Gaussian & $2.50$ & $6$   & $12\,000$  \\
Gaussian & $3.00$ & $6$   & $12\,000$  \\
Gaussian & $3.50$ & $6$   & $12\,000$  \\
Gaussian & $4.00$ & $5$   & $12\,000$  \\
Gaussian & $4.50$ & $5$   & $12\,000$  \\
Gaussian & $5.00$ & $5$   & $12\,000$  \\
Gaussian & $5.50$ & $5$   & $12\,000$  \\
Gaussian & $6.00$ & $4$   & $12\,000$  \\
Gaussian & $6.50$ & $4$   & $12\,000$  \\
Gaussian & $7.00$ & $4$   & $12\,000$  \\[2mm]
Bimodal  & $1.50$ & $6$   & $12\,000$  \\
Bimodal  & $2.00$ & $6$   & $12\,000$  \\
Bimodal  & $2.25$ & $6$   & $12\,000$  \\
Bimodal  & $2.50$ & $6$   & $12\,000$  \\
Bimodal  & $3.00$ & $6$   & $12\,000$  \\
Bimodal  & $3.50$ & $6$   & $12\,000$  \\
Bimodal  & $4.00$ & $6$   & $12\,000$  \\
Bimodal  & $4.50$ & $5$   & $12\,000$  \\
Bimodal  & $5.00$ & $5$   & $12\,000$  \\
\hline
\hline
\end{tabular*}
\end{table}

We measure the number of spins that flipped until the system regains
equilibrium and record the avalanche size distributions $D(n)$ for all
triggered avalanches of size $n$ until $S_i \rightarrow -S_i$ $\forall
i$.  When SOC is present (as for the SK model), we expect the avalanche
distributions to be power-law distributed with an exponential cutoff
that sets in at a characteristic size $n^*$.  Only if $n^*(N) \to
\infty$ for $N \to \infty$ without tuning any parameters does the system
exhibit true SOC.  $n^*$ is determined by fitting the tail of the
distributions to $D(n) \sim \exp[-n/n^*(N)]$ with $n^*(N)$ a fitting
parameter.  This procedure is repeated for different values of $\lambda$
and the thermodynamic value of $n*$ is determined by an extrapolation in
the system size $N$.

\subsection{Numerical results for Gaussian disorder} 

We start by showing avalanche distributions for selected values of the
exponent $\lambda$ which show the characteristic behavior of the system.

\begin{figure}
\includegraphics[width=\columnwidth]{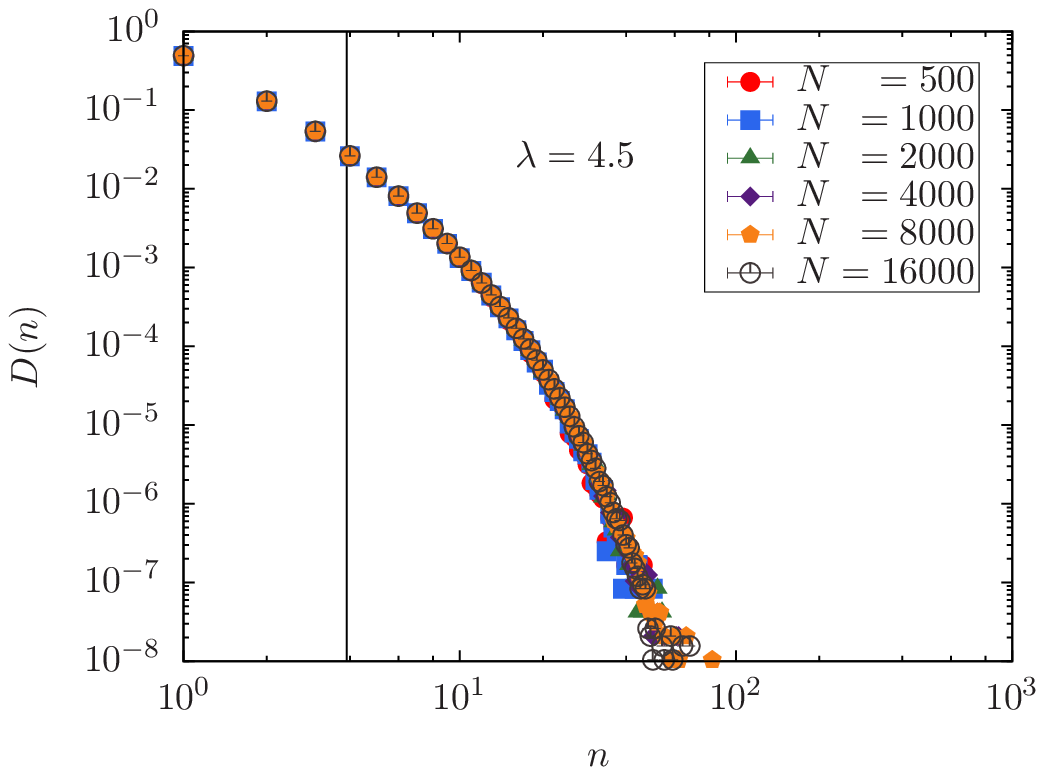}

\vspace*{1.0em}

\includegraphics[width=\columnwidth]{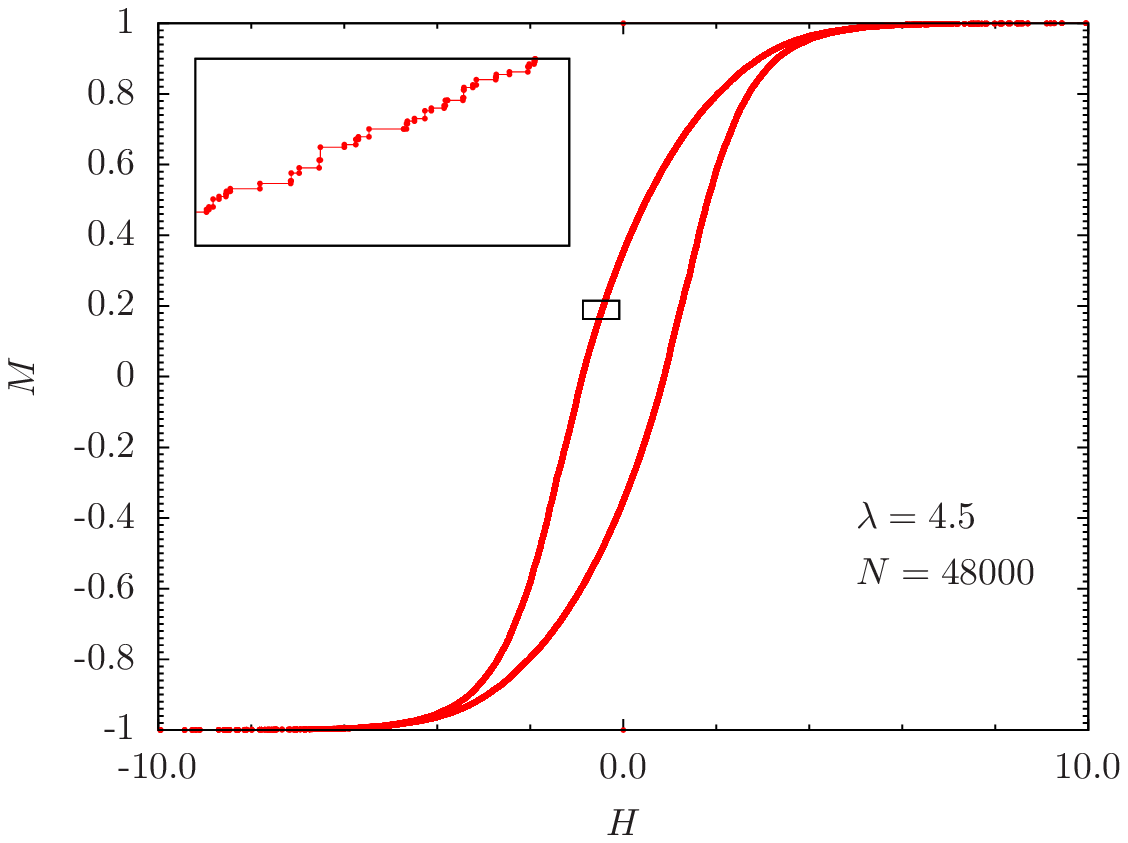}
\caption{(Color online)
Top: Avalanche distribution $D(n)$ for the Edwards-Anderson spin-glass
model with Gaussian disorder on scale-free networks with $\lambda=4.50$
recorded across the whole hysteresis loop. The data show no system size
dependence. The vertical (black) line marks the extrapolated value of
$n^*$. Clearly, no signs of SOC are visible in the data.  Bottom:
Magnetization $M = (1/N)\sum_i s_i$ versus field $H$ hysteresis loop for
$\lambda=4.50$ and $48000$ spins. The data are for one single sample and
meant as an illustration for the typical behavior of the system in a field.
The inset shows a zoom into the boxed region. The discrete steps due to
magnetization jumps in the hysteresis loop are clearly visible.
}
\label{fig:4.5}
\end{figure}

Figure \ref{fig:4.5} (top panel) shows avalanche distributions $D(n)$
for $\lambda=4.50$ recorded across the whole hysteresis loop (bottom
panel).  Here, the number of neighbors does not diverge with the system
size because $\lambda=4.50 > 2$. The distributions show no system size
dependence. The fact that the data show a curvature in a log-log plot
clearly indicate that these are not power laws. Although tens of
thousands of spins are simulated, the largest avalanches found span less
than 1\% of the system.  The vertical line represents the extrapolated
typical avalanche size $n^*$ which is rather small and indicates that
the system is not in an SOC state.

\begin{figure}
\includegraphics[width=\columnwidth]{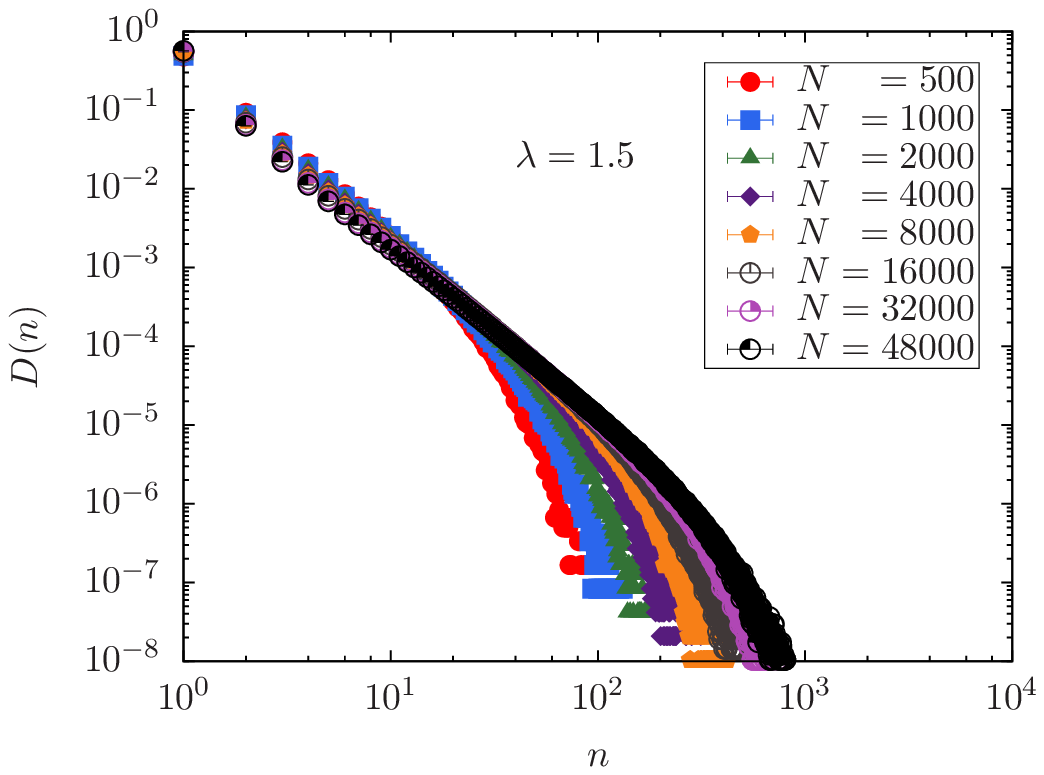}

\vspace*{1.0em}

\includegraphics[width=\columnwidth]{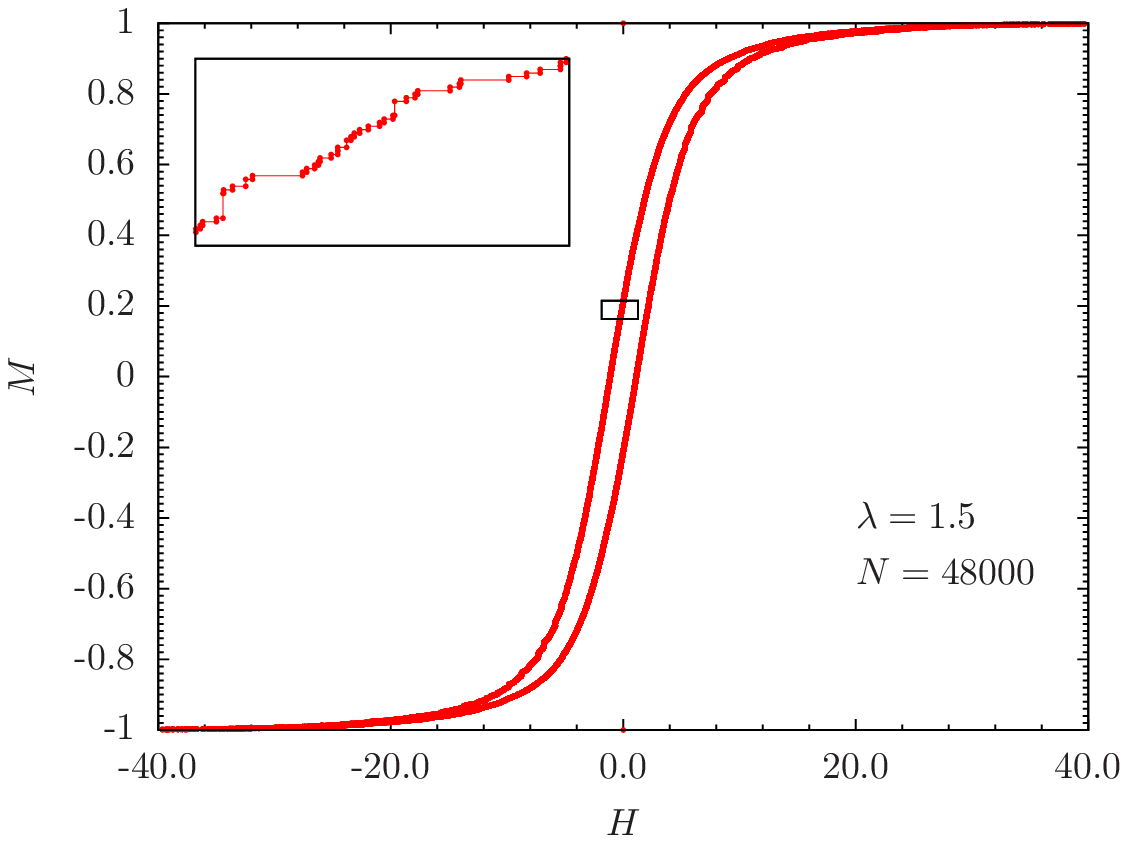}
\caption{(Color online)
Top: Avalanche distribution $D(n)$ for the Edwards-Anderson spin-glass
model with Gaussian disorder on scale-free networks with $\lambda=1.5$
recorded across the whole hysteresis loop. For $\lambda=1.5 < 2.0$ the
number of neighbors diverges.  The data show a clear system-size
dependence with the distributions becoming increasingly power-law-like
for increasing system size $N$.  As shown in Fig.~\ref{fig:soc_gauss},
the extrapolated cutoff value is $n^* = \infty$, i.e., the system
exhibits true SOC behavior.  Bottom: Magnetization $M = (1/N)\sum_i s_i$
versus field $H$ hysteresis loop for $\lambda=1.50$ and $48000$ spins. The
data are for one single sample and meant as an illustration for the typical 
behavior of the system in a field. The inset shows a zoom into the boxed
region. The discrete steps due to magnetization jumps in the hysteresis
loop are clearly visible. Qualitatively, the data seem to show larger
rearrangements as for $\lambda = 4.50$ (Fig.~\ref{fig:4.5}).
}
\label{fig:1.5}
\end{figure}

In contrast, Fig.~\ref{fig:1.5}, top panel, shows data for $\lambda=1.5
< 2$ in the regime where the number of neighbors diverges with the
system size.  The distributions $D(N)$ have a clearly visible power-law
behavior with a crossover size $n^*(N)$ that grows with increasing
system size.  Furthermore, a careful extrapolation to the thermodynamic
limit shows that $1/n^* = -0.0012(23)$, i.e., $n^* =\infty$.  The
hysteresis loop shown in the bottom panel of Fig.~\ref{fig:1.5} suggests
that for this value of $\lambda$ larger rearrangements of spins are
possible.

\begin{figure}
\includegraphics[width=\columnwidth]{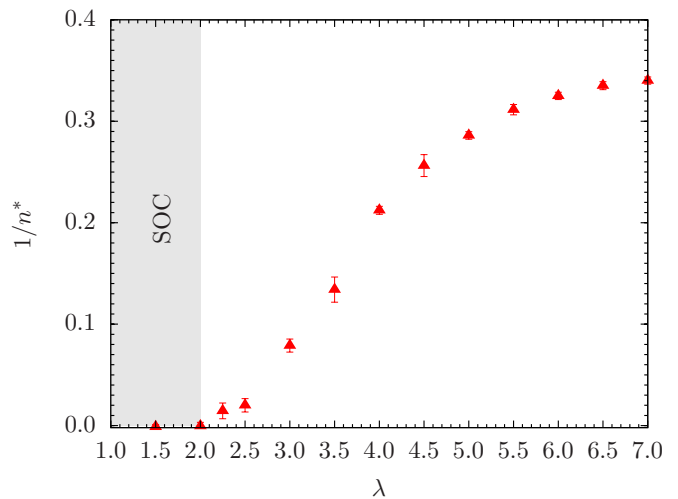}
\caption{(Color online)
Characteristic avalanche size $n^*$ extrapolated to the thermodynamic
limit for different values of $\lambda$ and Gaussian disorder. Plotted 
are $1/n^*$ versus $\lambda$. Only when $1/n^* = 0$ (here within error bars) 
we can expect the system to show SOC behavior. This is only the case 
for $\lambda \le 2$, i.e., in the regime where the number of neighbors 
diverges.
}
\label{fig:soc_gauss}
\end{figure}

We have repeated these simulations for several values of the exponent
$\lambda$. Our results are summarized in Fig.~\ref{fig:soc_gauss}, where
$1/n^*$ is plotted as a function of $\lambda$. Clearly, $1/n^* = 0$ only
if $\lambda \le 2$, i.e., in the regime where the number of neighbors
diverges, in perfect agreement with the results of
Ref.~\cite{andresen:13} for hypercubic systems, as well as the SK model
\cite{pazmandi:99}. Note that we have also recorded distributions of
magnetization jumps (not shown) \cite{pazmandi:99,andresen:13} that
qualitatively display the same behavior as the avalanche size
distributions.

\subsection{Numerical results for bimodal disorder}

So far, we have only probed for the existence of SOC within the
spin-glass phase. Bimodal disorder [Eq.~\eqref{eq:bim}] has the
advantage that one can easily tune the fraction of ferromagnetic bonds
by changing $p$. When $p = 1$ the system is a pure ferromagnet, whereas
for $p = 0$ it is an antiferromagnet and for $p = 0.5$ a spin glass
(comparable to the Gaussian case).

Sethna {\em et al.}, as well as others, have studied the random-field
Ising model
\cite{imry:75,sethna:93,vives:94,perkovic:95,perkovic:99,kuntz:98,sethna:04}
where the level of ferromagnetic behavior is tuned by changing the width
of the random-field distribution $\sigma$. In particular, for three
space dimensions, there is a critical value $\sigma_c$ where a jump in
the hysteresis loop appears, i.e., large system-spanning rearrangements
of the spins start to occur when $\sigma > \sigma_c$. We call this
regime {\em supercritical} because here system-spanning avalanches
will always occur in a predominant fashion. For $\sigma = \sigma_c$ true
power-law distributions of the spin avalanches are obtained, whereas for
$\sigma < \sigma_c$ no system-spanning rearrangements are found. We call
the latter scenario {\em subcritical}.

\begin{figure}
\includegraphics[width=\columnwidth]{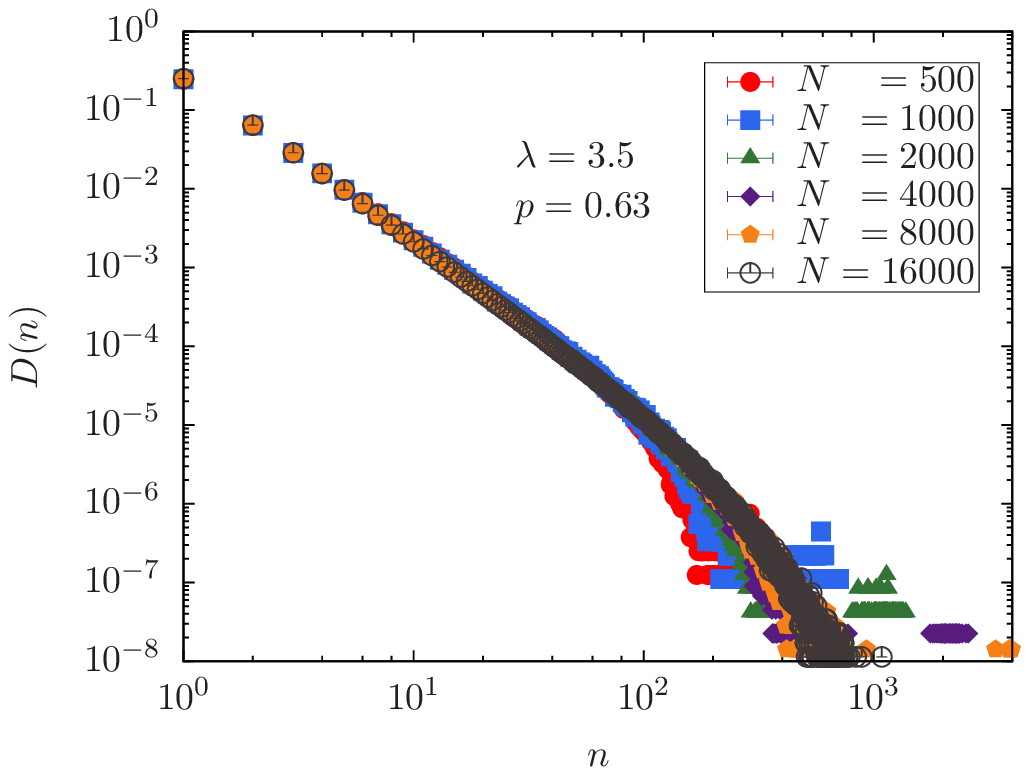}
\includegraphics[width=\columnwidth]{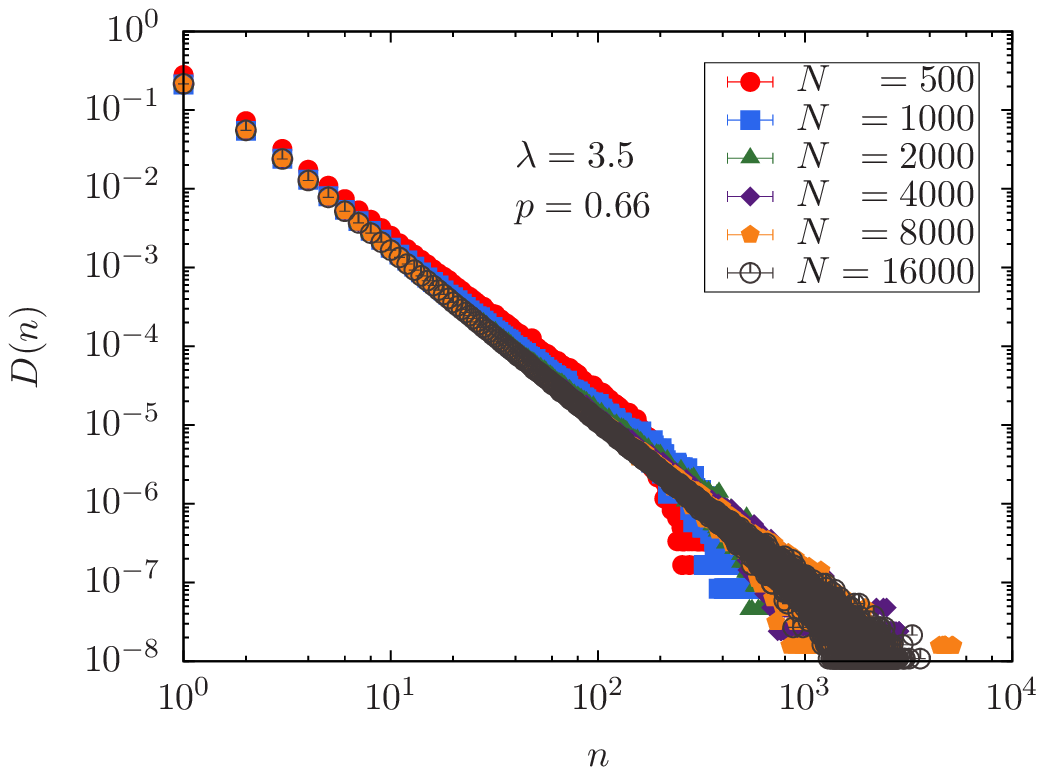}
\includegraphics[width=\columnwidth]{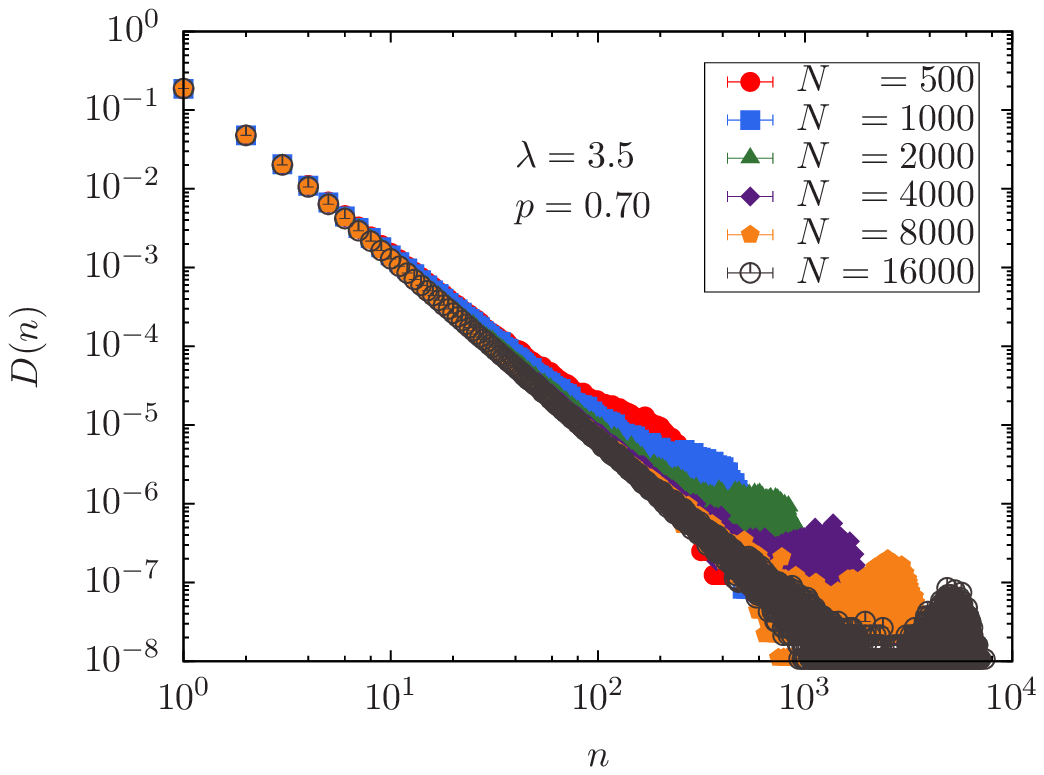}
\caption{(Color online)
Avalanche distribution $D(n)$ for the Edwards-Anderson spin-glass model
with bimodal disorder on scale-free networks with $\lambda=3.5$ recorded
across the whole hysteresis loop. Top panel: Data for $p = 0.63 < p_c$.
Here the system displays subcritical behavior, i.e., the characteristic
avalanche size $n^*$ is finite. Center panel: For $p = 0.66 \approx p_c$
the system is in the critical regime where the distributions are well
described by power laws. Bottom panel: For $p = 0.70 > p_c$ the system
is in the supercritical regime. A jump in the hysteresis loop occurs,
i.e., very large rearrangements are very probable, as can be seen in the
bump that develops in the distributions $D(n)$ for large $n$.
}
\label{fig:soc_critical}
\end{figure}

\begin{figure}
\includegraphics[width=\columnwidth]{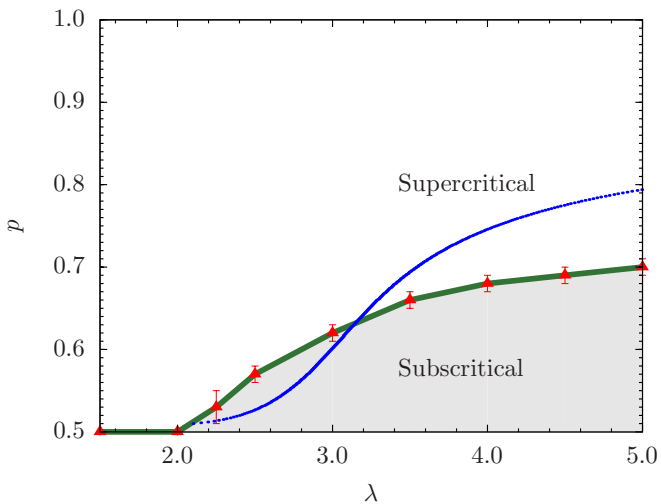}
\caption{(Color online)
Fraction of ferromagnetic bonds $p$ versus $\lambda$ phase diagram for the
Edwards-Anderson spin-glass model on scale free networks with bimodal
interactions between the spins. For $\lambda > 2$ a critical line
$p_c(\lambda)$ separates the subcritical regime where avalanches are
small, from the supercritical regime where system-spanning avalanches
are very common.  Along the critical line $p_c(\lambda)$ (triangles,
solid line) avalanche sizes are distributed according to power laws. For
$\lambda \le 2$ the number of neighbors diverges. In this regime for $p
= 0.5$ the system displays avalanches that are power laws, i.e., true
SOC.  The dotted line represents the spin-glass--to--ferromagnetic
phase boundary from Fig.~2 in Ref.~\cite{katzgraber:12}.
}
\label{fig:soc_bimodal}
\end{figure}

Here we find a similar behavior when tuning the fraction of
ferromagnetic bonds $p$.  Figure \ref{fig:soc_critical} shows the
typical behavior we observe for the avalanche distributions $D(n)$.  For
$p = 0.63$ and $\lambda = 3.50$ (Fig.~\ref{fig:soc_critical}, top panel),
the distributions show small system-size dependence. A detailed analysis
of the characteristic avalanche size $n^*(N)$ shows that it extrapolates
to a finite value in the thermodynamic limit. This means we are in the
subcritical regime.  However, for $\lambda = 3.50$ and $p = 0.66$ clear
power laws in the distributions $D(n)$ emerge
(Fig.~\ref{fig:soc_critical}, center panel).  Here $n^* \to \infty$,
i.e., true power-law behavior. However, for $\lambda = 3.50$ and $p =
0.70$, although most of the distributions show a clear power-law-like
behavior, a bump for large $n$ appears (Fig.~\ref{fig:soc_critical},
bottom panel). In this case the probability for very large
rearrangements increases. Direct inspection of the underlying hysteresis
loops (not shown) shows a jump in the magnetization, i.e., we are in the
supercritical regime.  We repeat these simulations for different
exponents $\lambda$ and vary the fraction of ferromagnetic bonds $p$
until the distributions are power laws. This allows us to construct the
phase diagram shown in Fig.~\ref{fig:soc_bimodal}. We find a critical
line $p_c(\lambda)$ (triangles, solid curve) that separates the
subcritical region from the supercritical region. Along the critical line
avalanche size distributions are power laws. Note that this critical
line shows no close correlations with the spin-glass--to--ferromagnetic
boundary computed in Ref.~\cite{katzgraber:12} (dotted line in
Fig.~\ref{fig:soc_bimodal}).  For $\lambda \le 2$ and when $p = 0.5$,
i.e., within the spin-glass phase where the graph connectivity diverges,
we recover true SOC.

\section{Summary and Conclusions}

We have studied Boolean (Ising) variables on a scale-free graph with
competing interactions in an external field both in thermal equilibrium,
as well as in a nonequilibrium hysteretic setting.

At finite temperatures, we show that for $\lambda > 3$, where at zero
field the system orders at finite temperatures \cite{katzgraber:12}, spin
glasses on scale-free graphs do order in a field, i.e., their behavior
is very much reminiscent of the mean-field SK model in a field. Naively,
one could have expected that outside the SK regime ($\lambda < 4$) a
behavior reminiscent of (diluted) one-dimensional spin glasses with
power-law interactions \cite{kotliar:83,katzgraber:03,leuzzi:08} emerges
where a spin-glass state in a field seems stable only within the
mean-field regime of the model \cite{katzgraber:05c,katzgraber:09b}.
These results again illustrate the superb robustness of Boolean decision
problems on scale-free networks to perturbations. In this case, a stable
spin-glass state emerges at nonzero temperatures even in the presence of
magnetic fields (external global biases).

At zero temperature, when driven with an external field, Boolean
decision problems on scale-free networks show self-organized critical
behavior only when the number of neighbors diverges with the system
size, i.e., for $\lambda \le 2$.  For $\lambda > 2$ and with bimodal
disorder, a behavior reminiscent of the random-field Ising model is
found
\cite{sethna:93,vives:94,perkovic:95,perkovic:99,kuntz:98,sethna:04}
where system-spanning avalanches only occur whenever the fraction of
ferromagnetic bonds $p_c(\lambda)$ is tuned towards a critical value.
These results show that ``damage'' can easily spread on real networks
where typically $\lambda \lesssim 3$. Therefore, in contrast the
robustness found at finite temperatures, Boolean decision problems on
scale-free networks show a potential fragility when driven in a
nonequilibrium scenario at zero temperature.

It will be interesting to perform these simulations for real networks in
the future, as well as the study of $q$-state Potts variables
\cite{yeomans:92}.

\begin{acknowledgments} 

We thank R.~S.~Andrist, V.~Dobrosavljevi{\'c}, K.~Janzen,
R.~B.~Macdonald, A.~P.~Young, and G.~T.~Zimanyi for fruitful
discussions.  In particular, we especially thank
C.~K.~Thomas for providing the code to produce the phase boundary in
Fig.~2 of Ref.~\cite{katzgraber:12}.  H.G.K.~also thanks
Privatbrauerei Franz Inselkammer for providing the necessary inspiration
for this project. We thank the NSF (Grant No.~DMR-1151387) for support
during government shutdown. Finally, we thank the Texas
Advanced Computing Center (TACC) at The University of Texas at Austin
for providing HPC resources (Lonestar Dell Linux Cluster) and Texas A\&M
University for access to their Eos cluster.

\end{acknowledgments}

\appendix* 

\section{Analytical form of the de Almeida-Thouless for $H_r \to 0$}
\label{app:at}

In this appendix we derive analytically the form of the AT line in the
limit when $H_r \to 0$ for a type of scale-free network which is very
convenient for analytical calculations, namely the static model used by
Kim {\em et al}.~\cite{kim:05}, whose procedures and equations we shall
closely follow. In this model the number of vertices $N$ is fixed. Each
vertex $i$ ($i=1,2,\ldots,N$) is given a weight $p_i$, where
\begin{equation} 
p_i=\frac{i^{-\mu}}{\zeta_N(\mu)} .
\label{pidef} 
\end{equation} 
where $\mu$ is related to $\lambda$ via $\lambda=1+1/\mu$, and
\begin{equation}
\zeta_N(\mu)\equiv \sum_{j=1}^Nj^{-\mu} \approx \frac{N^{1-\mu}}{1-\mu} .
\end{equation}
Only $\mu$ in the range $[0,1)$ (i.e., $\lambda>2$) will be discussed.
Two vertices $i$ and $j$ are selected with probabilities $p_i$ and $p_j$
and if $i \ne j$ they are connected with a single bond unless the pair
are already connected. The process is repeated $NK/2$ times. Then in
such a network, the probability that a given pair of vertices is not
connected by an edge is $1-f_{ij}=(1-2p_ip_j)^{NK/2} \approx
\exp(-NKp_ip_j)$, and the probability that they are connected by an edge
is $f_{ij}=NKp_ip_j$. This product form for $f_{ij}$ enabled Kim {\em et
al}.~\cite{kim:05} to proceed analytically. Note that here $K$ is the
mean degree of the scale-free network generated by this procedure.

We shall work in the paramagnetic phase where the spin glass is replica 
symmetric, i.e., where
\begin{equation}
q_{ab}=\sum_ip_i\langle s_i^{a} s_i^{b}\rangle =q,
\label{qdef}
\end{equation}
independently of the replica labels $a =1,2,\ldots ,n$, where $n$ is set
to zero at the end of the calculation. In $q_{ab}$, $a \ne b$. Kim {\em
et al}.~\cite{kim:05} showed then that the higher order parameters such
as $q_{abcd} =\sum_i p_i\langle s_i^{a} s_i^{b} s_i^{c} s_i^{d} \rangle$
can be neglected when $q$ is sufficiently small---that is, in the region
near $T_c$ studied in this appendix---and that a ``truncation''
approximation can be made for $q$
\begin{equation}
q=\int {\cal D} z\sum_{i=1}^Np_i\tanh^2
\left(z\sqrt{NK{\bf T}_2p_iq+H_r^2/T^2}\right),
\label{qeqn}
\end{equation}
where 
\begin{equation}
\int{\cal D}z\equiv \frac{1}{\sqrt{2\pi}}
\int_{-\infty}^{\infty}dze^{-z^2/2}
\end{equation}
and
\begin{equation}
{\bf T}_2
=\langle \tanh^2( J_{ij}/T)\rangle .
\end{equation}
Here the average is over the distribution of bonds, assumed symmetric,
i.e., $P(J_{ij})=P(-J_{ij})$. The random field of variance $H_r^2$ was
not included in the Kim {\em et al}.~\cite{kim:05} paper, but
Eq.~(\ref{qeqn}) is consistent with the equations for a spin glass in a
random field studied in Ref.~\cite{sharma:10} (in the appropriate
limit).

In the $H_r$--$T$ phase diagram it is expected that the assumption of
replica symmetry holds until the AT line is crossed. The equation of the
line where the spin-glass susceptibility diverges follows from the
expressions given in Ref.~\cite{kim:05}:
\begin{equation}
(K{\bf T}_2)^{-1}=\int \!\! {\cal D}z\sum_{i=1}^N Np_i^2 {\rm sech}^4
\left(z \sqrt{NK{\bf T}_2p_iq + H_r^2/T^2}\right).
\label{ATeqn}
\end{equation}
The solution of Eqs.~(\ref{qeqn}) and (\ref{ATeqn}) together fix the
equation of the AT line.

It is convenient to convert the sums over $i$ to integrals. Let $x=i/N$.
Then $\sum_{i=1}^N \to \int_0^1 N dx$, and in the large-$N$ limit
Eq.~(\ref{qeqn}) becomes
\begin{equation}
q =\int {\cal D}z \int_0^{1}dx \frac{1-\mu}{x^{\mu}}
\tanh^2\left(z\sqrt{Q^{\prime}/x^{\mu}+ H_r^2/T^2}\right) ,
\label{qxeqn}
\end{equation}
where $Q^{\prime}=(1-\mu)K{\bf T}_2q$. Equation (\ref{ATeqn}) becomes on
converting the sum to an integral
\begin{eqnarray}
(K{\bf T}_2)^{-1}&=&\int{\cal D}z\int_0^1dx \frac {(1-\mu)^2}{x^{2
\mu}}\nonumber \\&& \;\;\;\;\;\;\; {\rm sech}^4
\left(z\sqrt{Q^{\prime}/x^{\mu}+H_r^2/T^2}\right).
\label{ATxeqn}
\end{eqnarray}

We shall only study explicitly here the case where $3 < \lambda <4$
($1/3< \mu <1/2$). Similar procedures can be used to determine the AT
line when $\lambda > 4$. We first rewrite Eq.~(\ref{qxeqn}) as
\begin{widetext}
\begin{equation}
q =\int {\cal D}z \int_0^{1}dx
\frac{1-\mu}{x^{\mu}}\left\{z^2(Q^{\prime}/x^{\mu}+ H_r^2/T^2)
+\left[\tanh^2\left(z\sqrt{Q^{\prime}/x^{\mu}+ H_r^2/T^2}\right)-
z^2(Q^{\prime}/x^{\mu}+ H_r^2/T^2)\right]\right\}.
\label{qsplit}
\end{equation}
\end{widetext}
The integral over $z$ involving just the first line of
Eq.~(\ref{qsplit}) can be done to yield
\begin{equation}
q=(K/K_p) {\bf T}_2q+ H_r^2/T^2 +R(H_r,q) ,
\label{qR}
\end{equation} 
where 
\begin{equation}
K_p= \frac{1-2 \mu}{(1-\mu)^2},
\label{Kpdef}
\end{equation}
and
\begin{widetext}
\begin{equation}
R(H_r,q)=\int {\cal D}z \int_0^{1}dx \frac{1-\mu}{x^{\mu}}
\left[\tanh^2\left(z\sqrt{Q^{\prime}/x^{\mu}+ H_r^2/T^2})-
z^2(Q^{\prime}/x^{\mu}+ H_r^2/T^2\right)\right] .
\label{Rdef}
\end{equation}
\end{widetext}
One can show that $R(H_r,q)=R(0,q)+{\mathcal O}(Q^{\prime} H_r^2/T^2)$.
For small $q$, the term in addition to $R(0,q)$ is negligible in
comparison to the term $ H_r^2/T^2$ in Eq.~(\ref{qR}) and can be
dropped. We next re-write the integral for $R(0,q)$ as
\begin{widetext}
\begin{equation}
R(0,q)=\int {\cal D}z \left(\int_0^{\infty}dx-\int_1^{\infty}dx\right)\frac{1-\mu}{x^{\mu}}
  \big[\tanh^2(z\sqrt{Q^{\prime}/x^{\mu}})-
z^2 Q^{\prime}/x^{\mu}\big] .
\label{Rsubdef}
\end{equation}
\end{widetext}
The integral from $1$ to $\infty$ can be evaluated for small
$Q^{\prime}$ by expanding the $\tanh$ in a power series in $Q^{\prime}$.
The integrals converge for $\lambda <4$ and the leading contribution is
$$2 {Q^{\prime}}^2 (1-\mu) \frac{\lambda-1}{4-\lambda} + {\mathcal
O}({Q^{\prime}}^3).$$ The integral from $0$ to $\infty$ can by evaluated
after a variable change $w=z\sqrt{Q^{\prime}/x^{\mu}}$ when it gives a
contribution $F(\lambda) {Q^{\prime}}^{\lambda-2}$, where
\begin{equation}
F(\lambda)=P(\lambda)\int_0^{\infty} dw \, w^{3-2 \lambda}[\tanh^2w-w^2] .
\end{equation}
Here 
$P(\lambda)=(1-\mu)\Gamma(\lambda-3/2)2^{\lambda-1}(\lambda-1)/\sqrt{\pi}$.
Thus, for $3 < \lambda <4$, the equation of state is
\begin{eqnarray}
H_r^2/T^2&=&q[1-K{\bf T}_2/K_p]-F(\lambda) {Q^{\prime}}^{\lambda-2}\nonumber
\\&& \;\; - 2 {Q^{\prime}}^2 (1-\mu) \frac{\lambda-1}{4-\lambda} + {\mathcal
O}({Q^{\prime}}^3) ,
\label{eqn34}
\end{eqnarray}
which agrees with the expression given in Ref.~\cite{kim:05} when $
H_r=0$.

When $4 < \lambda < 5$, one can proceed in a similar fashion. The
equation of state is unchanged except $F(\lambda)$ becomes
$\tilde{F}(\lambda)$  where
\begin{equation}
\tilde{F}(\lambda)=P(\lambda)\int_0^{\infty} dw\, w^{3-2
\lambda}[\tanh^2w-w^2+2 w^4/3] .
\label{eqn45}
\end{equation}
For $\lambda >5$ the term in ${Q^{\prime}}^{\lambda-2}$ is subdominant
to the term of order ${Q^{\prime}}^3$ and can be ignored to leading
order.

We next deduce some simple features which follow from the equations of
state.  In the high-temperature state $q \sim  H_r^2/T^2$, and in the
limit of $ H_r/T \to 0$,
\begin{equation}
\chi_{SG} \to \frac{q}{(H_r^2/T^2)}=\frac{1}{1-K{\bf T}_2/K_p}.
\label{chiSG}
\end{equation}
The zero-field spin-glass susceptibility $\chi$ diverges at the
zero-field transition temperature $T_c$ where ${\bf T}_2=K_p/K$, and at
lower temperatures $q$ becomes nonzero. The divergence of this
susceptibility as the transition is approached is of the same form for
all $\lambda>3$. This means for the critical exponent
\begin{equation}
\gamma = 1 
\;\;\;\;\;\;\;\;\;\;\;\;\;\;\;\;\;\; 
\;\;\;\;\;\;\;\;\;\; 
(\lambda > 3)
\label{exp:gamma}
\end{equation}
However, the exponent $\beta$ in $q \sim (1-T/T_c)^{\beta}$ depends on
$\lambda$. We obtain
\begin{eqnarray}
\beta &=& \frac{1}{\lambda -3}
\;\;\;\;\;\;\;\;\;\; 
\;\;\;\;\;\;\; 
(3 < \lambda < 4) \\
\beta &=& 1
\;\;\;\;\;\;\;\;\;\; 
\;\;\;\;\;\;\;\;\;\; 
\;\;\;\; 
(\lambda > 4) .
\label{exp:beta}
\end{eqnarray}
We can use Eq.~(\ref{ATxeqn}) in conjunction with the equations of state
to determine the form of the AT line as $ H_r/T \to 0 $. Once again, we
shall start in the region $3 < \lambda <4$ and write the term ${\rm
sech}^4(z\sqrt{Q^{\prime}/x^{\mu}+ H_r^2/T^2})$ as $1+[{\rm
sech}^4(z\sqrt{Q^{\prime}/x^{\mu}+ H_r^2/T^2})-1]$. The term in unity in
the integral evaluates to $1/K_p$, so
\begin{equation} 
(K{\bf T}_2)^{-1}=1/K_p+S(H_r,q) ,
\label{AT34}
\end{equation}
where
\begin{widetext}
\begin{equation}
S(H_r,q)=\int{\cal D}z\int_0^1dx \frac {(1-\mu)^2}{x^{2 \mu}}
\left[{\rm sech}^4\left(z\sqrt{Q^{\prime}/x^{\mu}+ H_r^2/T^2}\right)-1\right].
\end{equation}
\end{widetext}
Once again, it is sufficient to evaluate $S(H_r,q)$ at $H_r=0$; the
corrections of ${\mathcal O}(H_r^2/T^2)$ are negligible compared to the
terms which we retain. Next we rewrite the integral as
\begin{widetext}
\begin{equation}
S(0,q)=\int {\cal D}z
\left(\int_0^{\infty}dx-\int_1^{\infty}dx\right)\frac{(1-\mu)^2}{x^{2\mu}} 
\left[{\rm sech}^4\left(z \sqrt{Q^{\prime}/x^{\mu}}\right)-1\right] .
\end{equation}
\end{widetext}
The integral from $0$ to $\infty$ can be evaluated after making the same
variable change $w=z\sqrt{Q^{\prime}/x^{\mu}}$, when it gives the contribution
$G(\lambda){Q^{\prime}}^{\lambda-3}$, where
\begin{widetext}
\begin{equation}
G(\lambda)= 2^{\lambda-2} (1-\mu)^2(\lambda-1) \Gamma(\lambda-5/2)/\sqrt{\pi}
\nonumber \\
\times \int_0^{\infty}dw \, w^{5-2\lambda}[{\rm sech}^4w-1].
\end{equation}
\end{widetext}
The integral from $1$ to $\infty$ can be done in a power series in
$Q^{\prime}$ and the leading term of this contribution to $S(0,q)$ is
\begin{equation}
-2(1-\mu)^2Q^{\prime}/(1-3\mu)+{\mathcal O}({Q^{\prime}}^2).
\nonumber
\end{equation}
We can now calculate the AT line: It is simplest to combine
Eqs.~(\ref{eqn34}) and (\ref{AT34}) to eliminate the term in $(1-K{\bf
T_2}/K_p)$ when one finds that
\begin{equation}
 H_{\rm AT}^2/T^2= C(\lambda) {Q^{\prime}}^{\lambda-2} +
{\mathcal O}({Q^{\prime}}^3),
\label{ATfin}
\end{equation}
where 
\begin{widetext}
\begin{equation}
C(\lambda)=\frac{1}{\sqrt{\pi}}2^{\lambda-2}(\lambda-2)\Gamma(\lambda-5/2)
\int_0^{\infty} dw \, w^{5-2 \lambda}\left\{{\rm sech}^4 w
-1 -2 (\lambda-5/2)\left[\tanh^2w/w^2-1\right] \right\} .
\label{Cdef}
\end{equation}
\end{widetext}
The integral has to be done numerically but it stays finite as $\lambda
\to 4$. For example, $C(3.75) \approx 0.530$. The terms of ${\mathcal
O}({Q^{\prime}}^2)$ cancel from Eq.~(\ref{ATfin}).  Thus, in the range $3
< \lambda < 4$, the equation of the AT line in terms of the temperature
rather than $Q^{\prime}$ is just
\begin{equation}
 H_{\rm AT}^2/T^2 \sim (1-T/T_c)^\frac{\lambda-2}{\lambda-3}
\;\;\;\;\;\;\;\;\;\; (3 < \lambda < 4) .
\label{hat34}
\end{equation}
Note that this is in agreement with the scaling form
\begin{equation}
H_{\rm AT}^2/T^2 \sim (1-T/T_c)^{\beta +\gamma},
\end{equation}
on inserting the vales for $\beta=1/(\lambda-3)$ and $\gamma=1$
for $3 < \lambda < 4$.

In the range $4<\lambda<5$, a similar expression holds for 
$H_{\rm AT}^2/T^2$ as in Eq.~(\ref{ATfin}), but $C(\lambda)$ becomes
$\tilde{C}(\lambda)$ where
\begin{widetext}
\begin{equation}
\tilde{C}(\lambda)=\frac{1}{\sqrt{\pi}}2^{\lambda-2}(\lambda-2)
\Gamma(\lambda-5/2)
\int_0^{\infty} dw \, w^{5-2 \lambda}\left\{{\rm sech}^4w-1+2 w^2
-2(\lambda-5/2)\left[\tanh^2w/w^2-1 +2 w^2/3\right]\right\} .
\label{Ctildedef}
\end{equation}
\end{widetext}
Because in this range the exponent $Q^{\prime} \sim (1-T/T_c)$, the form
of the AT line is
\begin{equation}
H_{\rm AT}^2/T^2 \sim (1-T/T_c)^{\lambda-2}
\;\;\;\;\;\;\;\;\;\; (4 < \lambda < 5) .
\label{hat45}
\end{equation}
Finally, in the range $\lambda >5$, the term in
${Q^{\prime}}^{\lambda-2}$ is subdominant compared with the term in
${Q^{\prime}}^3$ and
\begin{equation}
 H_{\rm AT}^2/T^2 \sim(1-T/T_c)^3
\;\;\;\;\;\;\;\;\;\; (\lambda > 5) ,
\label{hat5}
\end{equation}
which is the familiar form of the AT line in the SK model.

One can also use the static model to investigate the behavior when $\lambda
<3$. The spin-glass phase with broken replica symmetry exists in zero
field up to infinite temperature, i.e., $T_c$ is infinite when $\lambda
<3$ \cite{kim:05}. However, in the interval $5/2 < \lambda <3$ the
application of a large enough random field $H_r$ can restore replica
symmetry. By solving Eqs.~(\ref{qxeqn}) and (\ref{ATxeqn}) it can be
shown that this happens at a field $H_{\rm AT}$, where, as before,
$\beta^2 H_{\rm AT}^2 \sim {Q^{\prime}}^{\lambda-2}$ where
\begin{equation}
H_{\rm AT} \sim T^{\frac{5 - 2\lambda}{3 -\lambda}}
\;\;\;\;\;\;\;\;\;\; (2.5 < \lambda < 3) .
\label{2.5}
\end{equation}
for the limit when $T \to \infty$. This phase boundary is, as usual, for
the thermodynamic limit when $N \to \infty$. The behavior which would be
seen in simulations at finite system size $N$ will be complicated by an
unfamiliar finite-size behavior because, for this $\lambda$ range, $T_c$
at zero field is infinite. When $\lambda < 5/2$ we think that for all
$H_r$ and $T$ the spin-glass phase has broken replica symmetry and so as
a consequence, there will then be no AT line.

\bibliography{refs}

\begin{thebibliography}{50}
\expandafter\ifx\csname natexlab\endcsname\relax\def\natexlab#1{#1}\fi
\expandafter\ifx\csname bibnamefont\endcsname\relax
  \def\bibnamefont#1{#1}\fi
\expandafter\ifx\csname bibfnamefont\endcsname\relax
  \def\bibfnamefont#1{#1}\fi
\expandafter\ifx\csname citenamefont\endcsname\relax
  \def\citenamefont#1{#1}\fi
\expandafter\ifx\csname url\endcsname\relax
  \def\url#1{\texttt{#1}}\fi
\expandafter\ifx\csname urlprefix\endcsname\relax\def\urlprefix{URL }\fi
\providecommand{\bibinfo}[2]{#2}
\providecommand{\eprint}[2][]{\url{#2}}

\bibitem[{\citenamefont{Albert et~al.}(1999)\citenamefont{Albert, Jeong, and
  {Barab{\'a}si}}}]{albert:99}
\bibinfo{author}{\bibfnamefont{R.}~\bibnamefont{Albert}},
  \bibinfo{author}{\bibfnamefont{H.}~\bibnamefont{Jeong}}, \bibnamefont{and}
  \bibinfo{author}{\bibfnamefont{A.-L.} \bibnamefont{{Barab{\'a}si}}},
  \bibinfo{journal}{Nature} \textbf{\bibinfo{volume}{401}},
  \bibinfo{pages}{130} (\bibinfo{year}{1999}).

\bibitem[{\citenamefont{Bartolozzi et~al.}(2006)\citenamefont{Bartolozzi,
  Surungan, Leinweber, and Williams}}]{bartolozzi:06}
\bibinfo{author}{\bibfnamefont{M.}~\bibnamefont{Bartolozzi}},
  \bibinfo{author}{\bibfnamefont{T.}~\bibnamefont{Surungan}},
  \bibinfo{author}{\bibfnamefont{D.~B.} \bibnamefont{Leinweber}},
  \bibnamefont{and} \bibinfo{author}{\bibfnamefont{A.~G.}
  \bibnamefont{Williams}}, \bibinfo{journal}{Phys. Rev. B}
  \textbf{\bibinfo{volume}{73}}, \bibinfo{pages}{224419}
  (\bibinfo{year}{2006}).

\bibitem[{\citenamefont{Herrero}(2009)}]{herrero:09}
\bibinfo{author}{\bibfnamefont{C.~P.} \bibnamefont{Herrero}},
  \bibinfo{journal}{Eur. Phys. J. B} \textbf{\bibinfo{volume}{70}},
  \bibinfo{pages}{435} (\bibinfo{year}{2009}).

\bibitem[{\citenamefont{Lee et~al.}(2006)\citenamefont{Lee, Jeong, and
  Noh}}]{lee:06a}
\bibinfo{author}{\bibfnamefont{S.~H.} \bibnamefont{Lee}},
  \bibinfo{author}{\bibfnamefont{H.}~\bibnamefont{Jeong}}, \bibnamefont{and}
  \bibinfo{author}{\bibfnamefont{J.~D.} \bibnamefont{Noh}},
  \bibinfo{journal}{Phys. Rev. E} \textbf{\bibinfo{volume}{74}},
  \bibinfo{pages}{031118} (\bibinfo{year}{2006}).

\bibitem[{\citenamefont{Weigel and Johnston}(2007)}]{weigel:07}
\bibinfo{author}{\bibfnamefont{M.}~\bibnamefont{Weigel}} \bibnamefont{and}
  \bibinfo{author}{\bibfnamefont{D.}~\bibnamefont{Johnston}},
  \bibinfo{journal}{Phys. Rev. B} \textbf{\bibinfo{volume}{76}},
  \bibinfo{pages}{054408} (\bibinfo{year}{2007}).

\bibitem[{\citenamefont{Mooij and Kappen}(2004)}]{mooij:04}
\bibinfo{author}{\bibfnamefont{J.~M.} \bibnamefont{Mooij}} \bibnamefont{and}
  \bibinfo{author}{\bibfnamefont{H.~J.} \bibnamefont{Kappen}}
  (\bibinfo{year}{2004}), \bibinfo{note}{(arXiv:cond-mat/0408378)}.

\bibitem[{\citenamefont{Kim et~al.}(2005)\citenamefont{Kim, Rodgers, Kahng, and
  Kim}}]{kim:05}
\bibinfo{author}{\bibfnamefont{D.-H.} \bibnamefont{Kim}},
  \bibinfo{author}{\bibfnamefont{G.~J.} \bibnamefont{Rodgers}},
  \bibinfo{author}{\bibfnamefont{B.}~\bibnamefont{Kahng}}, \bibnamefont{and}
  \bibinfo{author}{\bibfnamefont{D.}~\bibnamefont{Kim}},
  \bibinfo{journal}{Phys. Rev. E} \textbf{\bibinfo{volume}{71}},
  \bibinfo{pages}{056115} (\bibinfo{year}{2005}).

\bibitem[{\citenamefont{Ferreira et~al.}(2010)\citenamefont{Ferreira, Mendes,
  and Ostilli}}]{ferreira:10}
\bibinfo{author}{\bibfnamefont{A.~L.} \bibnamefont{Ferreira}},
  \bibinfo{author}{\bibfnamefont{J.~F.~F.} \bibnamefont{Mendes}},
  \bibnamefont{and} \bibinfo{author}{\bibfnamefont{M.}~\bibnamefont{Ostilli}},
  \bibinfo{journal}{Phys. Rev. E} \textbf{\bibinfo{volume}{82}},
  \bibinfo{pages}{011141} (\bibinfo{year}{2010}).

\bibitem[{\citenamefont{Ostilli et~al.}(2011)\citenamefont{Ostilli, Ferreira,
  and Mendes}}]{ostilli:11}
\bibinfo{author}{\bibfnamefont{M.}~\bibnamefont{Ostilli}},
  \bibinfo{author}{\bibfnamefont{A.~L.} \bibnamefont{Ferreira}},
  \bibnamefont{and} \bibinfo{author}{\bibfnamefont{J.~F.~F.}
  \bibnamefont{Mendes}}, \bibinfo{journal}{Phys. Rev. E}
  \textbf{\bibinfo{volume}{83}}, \bibinfo{pages}{061149}
  (\bibinfo{year}{2011}).

\bibitem[{\citenamefont{{Katzgraber} et~al.}(2012)\citenamefont{{Katzgraber},
  {Janzen}, and {Thomas}}}]{katzgraber:12}
\bibinfo{author}{\bibfnamefont{H.~G.} \bibnamefont{{Katzgraber}}},
  \bibinfo{author}{\bibfnamefont{K.}~\bibnamefont{{Janzen}}}, \bibnamefont{and}
  \bibinfo{author}{\bibfnamefont{C.~K.} \bibnamefont{{Thomas}}},
  \bibinfo{journal}{Phys. Rev. E} \textbf{\bibinfo{volume}{86}},
  \bibinfo{pages}{031116} (\bibinfo{year}{2012}).

\bibitem[{\citenamefont{de~Almeida and Thouless}(1978)}]{almeida:78}
\bibinfo{author}{\bibfnamefont{J.~R.~L.} \bibnamefont{de~Almeida}}
  \bibnamefont{and} \bibinfo{author}{\bibfnamefont{D.~J.}
  \bibnamefont{Thouless}}, \bibinfo{journal}{J. Phys. A}
  \textbf{\bibinfo{volume}{11}}, \bibinfo{pages}{983} (\bibinfo{year}{1978}).

\bibitem[{\citenamefont{Sherrington and Kirkpatrick}(1975)}]{sherrington:75}
\bibinfo{author}{\bibfnamefont{D.}~\bibnamefont{Sherrington}} \bibnamefont{and}
  \bibinfo{author}{\bibfnamefont{S.}~\bibnamefont{Kirkpatrick}},
  \bibinfo{journal}{Phys. Rev. Lett.} \textbf{\bibinfo{volume}{35}},
  \bibinfo{pages}{1792} (\bibinfo{year}{1975}).

\bibitem[{\citenamefont{Newman and Stein}(1994)}]{newman:94}
\bibinfo{author}{\bibfnamefont{C.~M.} \bibnamefont{Newman}} \bibnamefont{and}
  \bibinfo{author}{\bibfnamefont{D.~L.} \bibnamefont{Stein}},
  \bibinfo{journal}{Phys. Rev. Lett.} \textbf{\bibinfo{volume}{72}},
  \bibinfo{pages}{2286} (\bibinfo{year}{1994}).

\bibitem[{\citenamefont{Cieplak et~al.}(1994)\citenamefont{Cieplak, Maritan,
  and Banavar}}]{cieplak:94}
\bibinfo{author}{\bibfnamefont{M.}~\bibnamefont{Cieplak}},
  \bibinfo{author}{\bibfnamefont{A.}~\bibnamefont{Maritan}}, \bibnamefont{and}
  \bibinfo{author}{\bibfnamefont{J.~R.} \bibnamefont{Banavar}},
  \bibinfo{journal}{Phys. Rev. Lett.} \textbf{\bibinfo{volume}{72}},
  \bibinfo{pages}{2320} (\bibinfo{year}{1994}).

\bibitem[{\citenamefont{Schenk et~al.}(2002)\citenamefont{Schenk, Drossel, and
  Schwabl}}]{schenk:02}
\bibinfo{author}{\bibfnamefont{K.}~\bibnamefont{Schenk}},
  \bibinfo{author}{\bibfnamefont{B.}~\bibnamefont{Drossel}}, \bibnamefont{and}
  \bibinfo{author}{\bibfnamefont{F.}~\bibnamefont{Schwabl}}, in
  \emph{\bibinfo{booktitle}{{Computational Statistical Physics}}}, edited by
  \bibinfo{editor}{\bibfnamefont{K.~H.} \bibnamefont{Hoffmann}}
  \bibnamefont{and} \bibinfo{editor}{\bibfnamefont{M.}~\bibnamefont{Schreiber}}
  (\bibinfo{publisher}{Springer-Verlag}, \bibinfo{address}{Berlin},
  \bibinfo{year}{2002}), p. \bibinfo{pages}{127}.

\bibitem[{\citenamefont{{P{\' a}zm{\' a}ndi} et~al.}(1999)\citenamefont{{P{\'
  a}zm{\' a}ndi}, {Zar{\' a}nd}, and {Zim{\' a}nyi}}}]{pazmandi:99}
\bibinfo{author}{\bibfnamefont{F.}~\bibnamefont{{P{\' a}zm{\' a}ndi}}},
  \bibinfo{author}{\bibfnamefont{G.}~\bibnamefont{{Zar{\' a}nd}}},
  \bibnamefont{and} \bibinfo{author}{\bibfnamefont{G.~T.} \bibnamefont{{Zim{\'
  a}nyi}}}, \bibinfo{journal}{Phys. Rev. Lett.} \textbf{\bibinfo{volume}{83}},
  \bibinfo{pages}{1034} (\bibinfo{year}{1999}).

\bibitem[{\citenamefont{Gon{\c c}alves and Boettcher}(2008)}]{goncalves:08}
\bibinfo{author}{\bibfnamefont{B.}~\bibnamefont{Gon{\c c}alves}}
  \bibnamefont{and}
  \bibinfo{author}{\bibfnamefont{S.}~\bibnamefont{Boettcher}},
  \bibinfo{journal}{J. Stat. Mech.}
  \textbf{\bibinfo{volume}{\normalfont{P01003}}} (\bibinfo{year}{2008}).

\bibitem[{\citenamefont{Andresen et~al.}(2013)\citenamefont{Andresen, Zhu,
  Andrist, Katzgraber, Dobrosavljevi{\'{c}}, and Zimanyi}}]{andresen:13}
\bibinfo{author}{\bibfnamefont{J.~C.} \bibnamefont{Andresen}},
  \bibinfo{author}{\bibfnamefont{Z.}~\bibnamefont{Zhu}},
  \bibinfo{author}{\bibfnamefont{R.~S.} \bibnamefont{Andrist}},
  \bibinfo{author}{\bibfnamefont{H.~G.} \bibnamefont{Katzgraber}},
  \bibinfo{author}{\bibfnamefont{V.}~\bibnamefont{Dobrosavljevi{\'{c}}}},
  \bibnamefont{and} \bibinfo{author}{\bibfnamefont{G.~T.}
  \bibnamefont{Zimanyi}}, \bibinfo{journal}{Phys. Rev. Lett.}
  \textbf{\bibinfo{volume}{111}}, \bibinfo{pages}{097203}
  (\bibinfo{year}{2013}).

\bibitem[{\citenamefont{{Sethna} et~al.}(1993)\citenamefont{{Sethna}, {Dahmen},
  {Kartha}, {Krumhansl}, {Roberts}, and {Shore}}}]{sethna:93}
\bibinfo{author}{\bibfnamefont{J.~P.} \bibnamefont{{Sethna}}},
  \bibinfo{author}{\bibfnamefont{K.}~\bibnamefont{{Dahmen}}},
  \bibinfo{author}{\bibfnamefont{S.}~\bibnamefont{{Kartha}}},
  \bibinfo{author}{\bibfnamefont{J.~A.} \bibnamefont{{Krumhansl}}},
  \bibinfo{author}{\bibfnamefont{B.~W.} \bibnamefont{{Roberts}}},
  \bibnamefont{and} \bibinfo{author}{\bibfnamefont{J.~D.}
  \bibnamefont{{Shore}}}, \bibinfo{journal}{Phys. Rev. Lett.}
  \textbf{\bibinfo{volume}{70}}, \bibinfo{pages}{3347} (\bibinfo{year}{1993}).

\bibitem[{\citenamefont{Perkovic et~al.}(1995)\citenamefont{Perkovic, Dahmen,
  and Sethna}}]{perkovic:95}
\bibinfo{author}{\bibfnamefont{O.}~\bibnamefont{Perkovic}},
  \bibinfo{author}{\bibfnamefont{K.~A.} \bibnamefont{Dahmen}},
  \bibnamefont{and} \bibinfo{author}{\bibfnamefont{J.~P.}
  \bibnamefont{Sethna}}, \bibinfo{journal}{Phys. Rev. Lett.}
  \textbf{\bibinfo{volume}{75}}, \bibinfo{pages}{4528} (\bibinfo{year}{1995}).

\bibitem[{\citenamefont{Perkovic et~al.}(1999)\citenamefont{Perkovic, Dahmen,
  and Sethna}}]{perkovic:99}
\bibinfo{author}{\bibfnamefont{O.}~\bibnamefont{Perkovic}},
  \bibinfo{author}{\bibfnamefont{K.~A.} \bibnamefont{Dahmen}},
  \bibnamefont{and} \bibinfo{author}{\bibfnamefont{J.~P.}
  \bibnamefont{Sethna}}, \bibinfo{journal}{Phys. Rev. B}
  \textbf{\bibinfo{volume}{59}}, \bibinfo{pages}{6106} (\bibinfo{year}{1999}).

\bibitem[{\citenamefont{Kuntz et~al.}(1998)\citenamefont{Kuntz, Perkovic,
  Dahmen, Roberts, and Sethna}}]{kuntz:98}
\bibinfo{author}{\bibfnamefont{M.~C.} \bibnamefont{Kuntz}},
  \bibinfo{author}{\bibfnamefont{O.}~\bibnamefont{Perkovic}},
  \bibinfo{author}{\bibfnamefont{K.~A.} \bibnamefont{Dahmen}},
  \bibinfo{author}{\bibfnamefont{B.~W.} \bibnamefont{Roberts}},
  \bibnamefont{and} \bibinfo{author}{\bibfnamefont{J.~P.} \bibnamefont{Sethna}}
  (\bibinfo{year}{1998}), \bibinfo{note}{(arXiv:cond-mat/9809122v2)}.

\bibitem[{\citenamefont{Sethna et~al.}(2004)\citenamefont{Sethna, Dahmen, and
  Perkovic}}]{sethna:04}
\bibinfo{author}{\bibfnamefont{J.~P.} \bibnamefont{Sethna}},
  \bibinfo{author}{\bibfnamefont{K.~A.} \bibnamefont{Dahmen}},
  \bibnamefont{and} \bibinfo{author}{\bibfnamefont{O.}~\bibnamefont{Perkovic}}
  (\bibinfo{year}{2004}), \bibinfo{note}{(arXiv:cond-mat/0406320v3)}.

\bibitem[{\citenamefont{{Vives} and {Planes}}(1994)}]{vives:94}
\bibinfo{author}{\bibfnamefont{E.}~\bibnamefont{{Vives}}} \bibnamefont{and}
  \bibinfo{author}{\bibfnamefont{A.}~\bibnamefont{{Planes}}},
  \bibinfo{journal}{Phys. Rev. B} \textbf{\bibinfo{volume}{50}},
  \bibinfo{pages}{3839} (\bibinfo{year}{1994}).

\bibitem[{\citenamefont{Vives and Planes}(2001)}]{vives:01}
\bibinfo{author}{\bibfnamefont{E.}~\bibnamefont{Vives}} \bibnamefont{and}
  \bibinfo{author}{\bibfnamefont{A.}~\bibnamefont{Planes}},
  \bibinfo{journal}{Phys. Rev. B} \textbf{\bibinfo{volume}{63}},
  \bibinfo{pages}{134431} (\bibinfo{year}{2001}).

\bibitem[{\citenamefont{Katzgraber et~al.}(2001)\citenamefont{Katzgraber,
  Palassini, and Young}}]{katzgraber:01}
\bibinfo{author}{\bibfnamefont{H.~G.} \bibnamefont{Katzgraber}},
  \bibinfo{author}{\bibfnamefont{M.}~\bibnamefont{Palassini}},
  \bibnamefont{and} \bibinfo{author}{\bibfnamefont{A.~P.} \bibnamefont{Young}},
  \bibinfo{journal}{Phys. Rev. B} \textbf{\bibinfo{volume}{63}},
  \bibinfo{pages}{184422} (\bibinfo{year}{2001}).

\bibitem[{\citenamefont{Katzgraber et~al.}(2009)\citenamefont{Katzgraber,
  Larson, and Young}}]{katzgraber:09b}
\bibinfo{author}{\bibfnamefont{H.~G.} \bibnamefont{Katzgraber}},
  \bibinfo{author}{\bibfnamefont{D.}~\bibnamefont{Larson}}, \bibnamefont{and}
  \bibinfo{author}{\bibfnamefont{A.~P.} \bibnamefont{Young}},
  \bibinfo{journal}{Phys. Rev. Lett.} \textbf{\bibinfo{volume}{102}},
  \bibinfo{pages}{177205} (\bibinfo{year}{2009}).

\bibitem[{\citenamefont{Barabasi and Albert}(1999)}]{barabasi:99}
\bibinfo{author}{\bibfnamefont{A.~L.} \bibnamefont{Barabasi}} \bibnamefont{and}
  \bibinfo{author}{\bibfnamefont{R.}~\bibnamefont{Albert}},
  \bibinfo{journal}{Science} \textbf{\bibinfo{volume}{286}},
  \bibinfo{pages}{509} (\bibinfo{year}{1999}).

\bibitem[{\citenamefont{Burda and Krzywicki}(2003)}]{burda:03}
\bibinfo{author}{\bibfnamefont{Z.}~\bibnamefont{Burda}} \bibnamefont{and}
  \bibinfo{author}{\bibfnamefont{A.}~\bibnamefont{Krzywicki}},
  \bibinfo{journal}{Phys. Rev. E} \textbf{\bibinfo{volume}{67}},
  \bibinfo{pages}{046118} (\bibinfo{year}{2003}).

\bibitem[{\citenamefont{Bogu\~{n}\'a et~al.}(2004)\citenamefont{Bogu\~{n}\'a,
  Pastor-Satorras, and Vespignani}}]{boguna:04}
\bibinfo{author}{\bibfnamefont{M.}~\bibnamefont{Bogu\~{n}\'a}},
  \bibinfo{author}{\bibfnamefont{R.}~\bibnamefont{Pastor-Satorras}},
  \bibnamefont{and}
  \bibinfo{author}{\bibfnamefont{A.}~\bibnamefont{Vespignani}},
  \bibinfo{journal}{Eur. Phys. J. B} \textbf{\bibinfo{volume}{38}},
  \bibinfo{pages}{205} (\bibinfo{year}{2004}).

\bibitem[{\citenamefont{Catanzaro et~al.}(2005)\citenamefont{Catanzaro,
  Bogu\~{n}\'{a}, and Pastor-Satorras}}]{catanzaro:05}
\bibinfo{author}{\bibfnamefont{M.}~\bibnamefont{Catanzaro}},
  \bibinfo{author}{\bibfnamefont{M.}~\bibnamefont{Bogu\~{n}\'{a}}},
  \bibnamefont{and}
  \bibinfo{author}{\bibfnamefont{R.}~\bibnamefont{Pastor-Satorras}},
  \bibinfo{journal}{Phys. Rev. E} \textbf{\bibinfo{volume}{71}},
  \bibinfo{pages}{027103} (\bibinfo{year}{2005}).

\bibitem[{\citenamefont{Young and Katzgraber}(2004)}]{young:04}
\bibinfo{author}{\bibfnamefont{A.~P.} \bibnamefont{Young}} \bibnamefont{and}
  \bibinfo{author}{\bibfnamefont{H.~G.} \bibnamefont{Katzgraber}},
  \bibinfo{journal}{Phys. Rev. Lett.} \textbf{\bibinfo{volume}{93}},
  \bibinfo{pages}{207203} (\bibinfo{year}{2004}).

\bibitem[{\citenamefont{Katzgraber and {Young}}(2005)}]{katzgraber:05c}
\bibinfo{author}{\bibfnamefont{H.~G.} \bibnamefont{Katzgraber}}
  \bibnamefont{and} \bibinfo{author}{\bibfnamefont{A.~P.}
  \bibnamefont{{Young}}}, \bibinfo{journal}{Phys. Rev. B}
  \textbf{\bibinfo{volume}{72}}, \bibinfo{pages}{184416}
  (\bibinfo{year}{2005}).

\bibitem[{\citenamefont{J{\"o}rg et~al.}(2008)\citenamefont{J{\"o}rg,
  Katzgraber, and Krzakala}}]{joerg:08a}
\bibinfo{author}{\bibfnamefont{T.}~\bibnamefont{J{\"o}rg}},
  \bibinfo{author}{\bibfnamefont{H.~G.} \bibnamefont{Katzgraber}},
  \bibnamefont{and} \bibinfo{author}{\bibfnamefont{F.}~\bibnamefont{Krzakala}},
  \bibinfo{journal}{Phys. Rev. Lett.} \textbf{\bibinfo{volume}{100}},
  \bibinfo{pages}{197202} (\bibinfo{year}{2008}).

\bibitem[{\citenamefont{{Ba{\~n}os} et~al.}(2012)\citenamefont{{Ba{\~n}os},
  {Cruz}, {Fernandez}, {Gil-Narvion}, {Gordillo-Guerrero}, {Guidetti},
  {I{\~n}iguez}, {Maiorano}, {Marinari}, {Martin-Mayor} et~al.}}]{banos:12}
\bibinfo{author}{\bibfnamefont{R.~A.} \bibnamefont{{Ba{\~n}os}}},
  \bibinfo{author}{\bibfnamefont{A.}~\bibnamefont{{Cruz}}},
  \bibinfo{author}{\bibfnamefont{L.~A.} \bibnamefont{{Fernandez}}},
  \bibinfo{author}{\bibfnamefont{J.~M.} \bibnamefont{{Gil-Narvion}}},
  \bibinfo{author}{\bibfnamefont{A.}~\bibnamefont{{Gordillo-Guerrero}}},
  \bibinfo{author}{\bibfnamefont{M.}~\bibnamefont{{Guidetti}}},
  \bibinfo{author}{\bibfnamefont{D.}~\bibnamefont{{I{\~n}iguez}}},
  \bibinfo{author}{\bibfnamefont{A.}~\bibnamefont{{Maiorano}}},
  \bibinfo{author}{\bibfnamefont{E.}~\bibnamefont{{Marinari}}},
  \bibinfo{author}{\bibfnamefont{V.}~\bibnamefont{{Martin-Mayor}}},
  \bibnamefont{et~al.}, \bibinfo{journal}{Proc. Natl. Acad. Sci. U.S.A.}
  \textbf{\bibinfo{volume}{109}}, \bibinfo{pages}{6452} (\bibinfo{year}{2012}).

\bibitem[{\citenamefont{{Baity-Jesi} et~al.}(2013)\citenamefont{{Baity-Jesi},
  {Alvarez Ba{\~n}os}, {Cruz}, {Fernandez}, {Gil-Narvion}, {Gordillo-Guerrero},
  {I{\~n}iguez}, {Maiorano}, {Mantovani}, {Marinari} et~al.}}]{baity:13}
\bibinfo{author}{\bibfnamefont{M.}~\bibnamefont{{Baity-Jesi}}},
  \bibinfo{author}{\bibfnamefont{R.}~\bibnamefont{{Alvarez Ba{\~n}os}}},
  \bibinfo{author}{\bibfnamefont{A.}~\bibnamefont{{Cruz}}},
  \bibinfo{author}{\bibfnamefont{L.~A.} \bibnamefont{{Fernandez}}},
  \bibinfo{author}{\bibfnamefont{J.~M.} \bibnamefont{{Gil-Narvion}}},
  \bibinfo{author}{\bibnamefont{{Gordillo-Guerrero}}},
  \bibinfo{author}{\bibfnamefont{D.}~\bibnamefont{{I{\~n}iguez}}},
  \bibinfo{author}{\bibfnamefont{A.}~\bibnamefont{{Maiorano}}},
  \bibinfo{author}{\bibfnamefont{F.}~\bibnamefont{{Mantovani}}},
  \bibinfo{author}{\bibfnamefont{E.}~\bibnamefont{{Marinari}}},
  \bibnamefont{et~al.} (\bibinfo{year}{2013}),
  \bibinfo{note}{(arxiv:cond-mat/1307.4998)}.

\bibitem[{\citenamefont{Binder}(1981)}]{binder:81}
\bibinfo{author}{\bibfnamefont{K.}~\bibnamefont{Binder}},
  \bibinfo{journal}{Phys. Rev. Lett.} \textbf{\bibinfo{volume}{47}},
  \bibinfo{pages}{693} (\bibinfo{year}{1981}).

\bibitem[{\citenamefont{Ciria et~al.}(1993)\citenamefont{Ciria, Parisi, Ritort,
  and Ruiz-Lorenzo}}]{ciria:93b}
\bibinfo{author}{\bibfnamefont{J.~C.} \bibnamefont{Ciria}},
  \bibinfo{author}{\bibfnamefont{G.}~\bibnamefont{Parisi}},
  \bibinfo{author}{\bibfnamefont{F.}~\bibnamefont{Ritort}}, \bibnamefont{and}
  \bibinfo{author}{\bibfnamefont{J.~J.} \bibnamefont{Ruiz-Lorenzo}},
  \bibinfo{journal}{J. Phys. I France} \textbf{\bibinfo{volume}{3}},
  \bibinfo{pages}{2207} (\bibinfo{year}{1993}).

\bibitem[{\citenamefont{Larson et~al.}(2013)\citenamefont{Larson, Katzgraber,
  Moore, and Young}}]{larson:13}
\bibinfo{author}{\bibfnamefont{D.}~\bibnamefont{Larson}},
  \bibinfo{author}{\bibfnamefont{H.~G.} \bibnamefont{Katzgraber}},
  \bibinfo{author}{\bibfnamefont{M.~A.} \bibnamefont{Moore}}, \bibnamefont{and}
  \bibinfo{author}{\bibfnamefont{A.~P.} \bibnamefont{Young}},
  \bibinfo{journal}{Phys. Rev. B} \textbf{\bibinfo{volume}{87}},
  \bibinfo{pages}{024414} (\bibinfo{year}{2013}).

\bibitem[{\citenamefont{Geyer}(1991)}]{geyer:91}
\bibinfo{author}{\bibfnamefont{C.}~\bibnamefont{Geyer}}, in
  \emph{\bibinfo{booktitle}{23rd Symposium on the Interface}}, edited by
  \bibinfo{editor}{\bibfnamefont{E.~M.} \bibnamefont{Keramidas}}
  (\bibinfo{publisher}{Interface Foundation}, \bibinfo{address}{Fairfax
  Station, VA}, \bibinfo{year}{1991}), p. \bibinfo{pages}{156}.

\bibitem[{\citenamefont{Hukushima and Nemoto}(1996)}]{hukushima:96}
\bibinfo{author}{\bibfnamefont{K.}~\bibnamefont{Hukushima}} \bibnamefont{and}
  \bibinfo{author}{\bibfnamefont{K.}~\bibnamefont{Nemoto}},
  \bibinfo{journal}{J. Phys. Soc. Jpn.} \textbf{\bibinfo{volume}{65}},
  \bibinfo{pages}{1604} (\bibinfo{year}{1996}).

\bibitem[{\citenamefont{Katzgraber et~al.}(2006)\citenamefont{Katzgraber,
  K\"orner, and Young}}]{katzgraber:06}
\bibinfo{author}{\bibfnamefont{H.~G.} \bibnamefont{Katzgraber}},
  \bibinfo{author}{\bibfnamefont{M.}~\bibnamefont{K\"orner}}, \bibnamefont{and}
  \bibinfo{author}{\bibfnamefont{A.~P.} \bibnamefont{Young}},
  \bibinfo{journal}{Phys. Rev. B} \textbf{\bibinfo{volume}{73}},
  \bibinfo{pages}{224432} (\bibinfo{year}{2006}).

\bibitem[{\citenamefont{Binder and Young}(1986)}]{binder:86}
\bibinfo{author}{\bibfnamefont{K.}~\bibnamefont{Binder}} \bibnamefont{and}
  \bibinfo{author}{\bibfnamefont{A.~P.} \bibnamefont{Young}},
  \bibinfo{journal}{Rev. Mod. Phys.} \textbf{\bibinfo{volume}{58}},
  \bibinfo{pages}{801} (\bibinfo{year}{1986}).

\bibitem[{\citenamefont{Katzgraber et~al.}(2002)\citenamefont{Katzgraber, {P{\'
  a}zm{\' a}ndi}, {Pike}, {Liu}, {Scalettar}, {Verosub}, and {Zim{\'
  a}nyi}}}]{katzgraber:02b}
\bibinfo{author}{\bibfnamefont{H.~G.} \bibnamefont{Katzgraber}},
  \bibinfo{author}{\bibfnamefont{F.}~\bibnamefont{{P{\' a}zm{\' a}ndi}}},
  \bibinfo{author}{\bibfnamefont{C.~R.} \bibnamefont{{Pike}}},
  \bibinfo{author}{\bibfnamefont{K.}~\bibnamefont{{Liu}}},
  \bibinfo{author}{\bibfnamefont{R.~T.} \bibnamefont{{Scalettar}}},
  \bibinfo{author}{\bibfnamefont{K.~L.} \bibnamefont{{Verosub}}},
  \bibnamefont{and} \bibinfo{author}{\bibfnamefont{G.~T.} \bibnamefont{{Zim{\'
  a}nyi}}}, \bibinfo{journal}{Phys. Rev. Lett.} \textbf{\bibinfo{volume}{89}},
  \bibinfo{pages}{257202} (\bibinfo{year}{2002}).

\bibitem[{\citenamefont{{Imry} and {Ma}}(1975)}]{imry:75}
\bibinfo{author}{\bibfnamefont{Y.}~\bibnamefont{{Imry}}} \bibnamefont{and}
  \bibinfo{author}{\bibfnamefont{S.-K.} \bibnamefont{{Ma}}},
  \bibinfo{journal}{Phys. Rev. Lett.} \textbf{\bibinfo{volume}{35}},
  \bibinfo{pages}{1399} (\bibinfo{year}{1975}).

\bibitem[{\citenamefont{{Kotliar} et~al.}(1983)\citenamefont{{Kotliar},
  {Anderson}, and {Stein}}}]{kotliar:83}
\bibinfo{author}{\bibfnamefont{G.}~\bibnamefont{{Kotliar}}},
  \bibinfo{author}{\bibfnamefont{P.~W.} \bibnamefont{{Anderson}}},
  \bibnamefont{and} \bibinfo{author}{\bibfnamefont{D.~L.}
  \bibnamefont{{Stein}}}, \bibinfo{journal}{Phys. Rev. B}
  \textbf{\bibinfo{volume}{27}}, \bibinfo{pages}{602} (\bibinfo{year}{1983}).

\bibitem[{\citenamefont{Katzgraber and Young}(2003)}]{katzgraber:03}
\bibinfo{author}{\bibfnamefont{H.~G.} \bibnamefont{Katzgraber}}
  \bibnamefont{and} \bibinfo{author}{\bibfnamefont{A.~P.} \bibnamefont{Young}},
  \bibinfo{journal}{Phys. Rev. B} \textbf{\bibinfo{volume}{67}},
  \bibinfo{pages}{134410} (\bibinfo{year}{2003}).

\bibitem[{\citenamefont{{Leuzzi} et~al.}(2008)\citenamefont{{Leuzzi}, {Parisi},
  {Ricci-Tersenghi}, and {Ruiz-Lorenzo}}}]{leuzzi:08}
\bibinfo{author}{\bibfnamefont{L.}~\bibnamefont{{Leuzzi}}},
  \bibinfo{author}{\bibfnamefont{G.}~\bibnamefont{{Parisi}}},
  \bibinfo{author}{\bibfnamefont{F.}~\bibnamefont{{Ricci-Tersenghi}}},
  \bibnamefont{and} \bibinfo{author}{\bibfnamefont{J.~J.}
  \bibnamefont{{Ruiz-Lorenzo}}}, \bibinfo{journal}{Phys. Rev. Lett.}
  \textbf{\bibinfo{volume}{101}}, \bibinfo{pages}{107203}
  (\bibinfo{year}{2008}).

\bibitem[{\citenamefont{Yeomans}(1992)}]{yeomans:92}
\bibinfo{author}{\bibfnamefont{J.~M.} \bibnamefont{Yeomans}},
  \emph{\bibinfo{title}{{Statistical Mechanics of Phase Transitions}}}
  (\bibinfo{publisher}{Oxford University Press}, \bibinfo{address}{Oxford},
  \bibinfo{year}{1992}).

\bibitem[{\citenamefont{Sharma and Young}(2010)}]{sharma:10}
\bibinfo{author}{\bibfnamefont{A.}~\bibnamefont{Sharma}} \bibnamefont{and}
  \bibinfo{author}{\bibfnamefont{A.~P.} \bibnamefont{Young}},
  \bibinfo{journal}{Phys. Rev. E} \textbf{\bibinfo{volume}{81}},
  \bibinfo{pages}{061115} (\bibinfo{year}{2010}).

\end{thebibliography}

\end{document}